\documentstyle[graphics,psfig,mysty]{mn}

\title[Modelling the Lyman lines]{Modelling the H Lyman lines in evolved late-type stars}
\author[S. A. Sim] {S. A. Sim\footnotemark \\ Department of Physics
(Theoretical Physics), University of Oxford, 1 Keble Road, Oxford, OX1 3NP,
UK}

\date{26 April 2001}

\pagerange{\pageref{firstpage}--\pageref{lastpage}} \pubyear{2001}

\begin{document}
\maketitle
\label{firstpage}

\begin{abstract}
The importance of Partial Redistribution (PRD) in the modelling of the
Lyman~$\alpha$ and Lyman~$\beta$ emission lines of hydrogen in stellar 
atmospheres is examined
using simple atmospheric models of a range of late-type stars. These
models represent the sub-giant Procyon (F5~IV-V), and the two giants
$\beta$~Gem (K0~III) and $\alpha$~Tau (K5~III). These stars are
selected to span a wide range of surface gravities: $1.25 < \log g <
4.00$. The calculations are performed using the computer code MULTI
\myshortcite{MULTI1} with the modifications made by \mylongcite{HL}.  It
is found that PRD effects are highly significant, both in the direct
prediction of the Lyman line profiles and in the application of
hydrostatic equilibrium (HSE) to calculate the atmospheric electron
density in static atmospheric models.
\end{abstract}


\section{Introduction}

\footnotetext{E-mail: sim@thphys.ox.ac.uk}

The modelling of the Lyman lines of hydrogen is extremely important
for the study of stellar atmospheres. The Lyman~$\alpha$ line is the
strongest single line in the ultraviolet spectra of cool stars and
represents a significant energy loss mechanism from the high
chromosphere and low transition region in most stars. The strength and
width of the Lyman lines, particularly Lyman~$\alpha$ and
Lyman~$\beta$, is such that they are important sources for
fluorescence in the lines of other elements \myshortcite{Jordan84}. This
fluorescence is of particular importance in low gravity stars such as
$\alpha$~Tau (McMurry, Jordan \& Carpenter \myfirstcite{Mcmurry2}).

Unfortunately, the Lyman lines are difficult to model since the
extreme optical thickness of the lines and the low electron densities in
the region of line formation makes the treatment of photon scattering
very important. The assumption of complete (frequency) redistribution
(CRD), which is often used in the modelling of the lines of other
atomic species, is unacceptable because most scattering events in a
transition such as Lyman~$\alpha$ are coherent (in a giant star such
as $\beta$~Gem in a region where the electron 
temperature $T_{e} \sim 10^{4}$K and the electron 
density $n_{e} \sim 10^{9}$cm$^{-3}$, only $\sim 10^{-4}$ of the 
scattering events are
incoherent). This coherence leads to a correlation between the
absorbed and emitted frequencies in a scattering event and so
introduces an explicit coupling of the radiation field at different
frequencies. This significantly complicates the problem that is to be
solved. Thus scattering in these lines must be modelled using the more
complete theory of partial (frequency) redistribution (PRD).

\mylongcite{Linsky} gives a review of the work on PRD in stellar
atmospheres.  Early work on the Lyman~$\alpha$ line in the solar
chromosphere gave clear indications that PRD effects would be very
important \myshortcite{Milkey}. However, the treatment of
\mylongcite{Milkey} is not entirely satisfactory because it is based on
the redistribution function theory 
of Omont, Smith \& Cooper (\myfirstcite{OSC}). This
redistribution function is widely used in the study of PRD in lines of
Ca~{\sc ii} and Mg~{\sc ii}, but is not suitable for hydrogen because it 
assumes
that the coherence fraction is frequency independent. However, the
incoherence fraction is only frequency independent in the so-called
{\it Impact Limit} of collisional broadening which is valid only in
the very core of Lyman~$\alpha$. An adequate treatment of the line
wings requires a more sophisticated atomic theory (under the
conditions that exist in a giant star such as $\beta$~Gem the {\it
Impact Limit} is only valid for wavelength differences from line centre
of $\leq 10^{-4}$~\AA). Such a theory exists in the form of the
Unified Theory of Collisions (Smith, Cooper \& Vidal
\myfirstcite{Unify}). This theory is developed by Vidal, Cooper \& Smith
(\myfirstcite{JQSRT1}) and specifically used to study hydrogen in
stellar atmospheres by \mylongcite{Yelnik} and Cooper, Ballagh \& Hubeny
(\myfirstcite{CBH}). An alternative theory of collisional broadening is
discussed by Smith, Cooper \& Roszman (\myfirstcite{JQSRT3}). 

The application of the Unified Theory in the study of PRD in hydrogen
has been carried out by several authors. \mylongcite{Rocket} used a
Unified Theory treatment of both ion and electron collisions to
compute a Lyman~$\alpha$ profile to compare with solar
observations. Their calculations were based on the solar model of
Vernazza, Avrett \& Loeser (\myfirstcite{VALold}) which they modified to
obtain close agreement with the observed profile. In their extensive
study of solar chromospheric modelling, Vernazza, Avrett \& Loeser
(\myfirstcite{VAL}) also performed PRD calculations for hydrogen. In
that paper, the authors experimented by setting different maximum
coherence fractions and they found that enforcing a maximum coherence
fraction of 98 per cent gave the best agreement with the observed
data. However, they gave no physical reason why the incoherence
fraction should be larger than the theoretical
prediction. \mylongcite{LJ92} use a PRD treatment of the Lyman~$\alpha$
line when considering their joint photospheric/chromospheric models of
the stars g Her and TX Psc. They find that a PRD treatment of hydrogen
(compared to a CRD treatment) makes a significant difference to the
proton density in certain parts of their atmospheric models and is
therefore significant to the ionization balance in 
hydrogen.  \mylongcite{Ubk} introduced PRD to the radiative 
transfer code MULTI, which in its original
form is discussed by \mylongcite{MULTI1} 
and \mylongcite{MULTI2}. 
The treatment of PRD used by \mylongcite{Ubk} 
was based on {\it Impact Theory} and is suitable for
the Ca~{\sc ii} H and K lines but not for hydrogen. More recently, 
\mylongcite{HL} 
modified the MULTI code extensively so that it could include the
frequency dependent coherence fraction required for hydrogen. They
discussed the modifications to the code and the theory behind them and 
also successfully applied their code to the solar
chromospheric model C of \mylongcite{VAL}. 

The purpose of this paper is to present the results of applying the
version of MULTI written by \mylongcite{HL} to models of stars other
than the Sun, so that PRD effects can be investigated in lower surface
gravity and their importance to the modelling techniques can be
studied. Several limitations remain in the approach used by
\mylongcite{HL} and these remain in the work here. The primary
limitation of the computer code is that it implements an angle
averaged redistribution function. This means that it can only be
applied realistically to a static atmospheric model and so all the
models discussed here are static and contain no net flows. There is
also a minor assumption, made for computational ease, that when
calculating the collisional broadening, it is assumed that the proton
density is equal to the electron density. In a typical high
chromosphere/transition region this assumption is reasonable and so
should not be a source of significant error.

\section{Theory}
\label{theory} 

The detailed theory of PRD in hydrogen is discussed by
\mylongcite{CBH}. The theory they describe, which is that implemented by
\mylongcite{HL}, includes an acceptable treatment of resonant scattering
in both Lyman~$\alpha$ and Lyman~$\beta$ and also includes
cross-redistribution from Balmer~$\alpha$ to Lyman~$\beta$. The
details of the method are discussed by \mylongcite{HL}. A brief summary
of those parts of the theory which are of particular relevance are
given below.

The study of PRD requires the calculation of the redistribution
function $R$ which gives the probability that a photon of frequency
$\nu^{\prime}$ will be absorbed and re-emitted as a photon of
frequency $\nu$. It is convenient to work in the rest frame of the
absorbing atom (on the understanding that, at a later stage an average
over a Maxwellian velocity distribution of such atoms will be
performed).  Thus the required quantity is

\[
f( \nu^{\prime} ) p(\nu^{\prime}, \nu)
\]
where $f$ is the absorption profile (which describes the probability
that a photon of frequency $\nu^{\prime}$ is absorbed) and $p$ is the
probability that, given an absorption at frequency $\nu^{\prime}$
there is a subsequent re-emission at frequency $\nu$.

The quantities $f$ and $p$ can be calculated from the quantum
mechanical theory of radiation. As an example, for Lyman~$\alpha$ they
are given by

\begin{equation}
f_{1s2p}(\Delta \omega) = \frac{1}{\pi} \frac{ \Gamma_{2p} (\Delta
\omega)}{{\Delta \omega}^{2} + \Gamma_{2p} (\Delta \omega)^{2}}
\end{equation}
and
\begin{eqnarray}
p_{1s2p}(\Delta \omega^{\prime}, \Delta \omega) & = &  (1-
\Lambda_{1s2p} (\Delta \omega^{\prime})) \; \delta( \Delta \omega -
\Delta \omega^{\prime} ) \nonumber \\ & & + \Lambda_{1s2p}(\Delta
\omega^{\prime}) f_{1s2p}(\Delta \omega)
\end{eqnarray}
where $\Delta \omega$ and $\Delta \omega^{\prime}$ are the angular
frequencies (measured relative to line centre) of the emitted and
absorbed photons, $\Gamma_{2p}$ is the total half width of the 2p
state and $\Lambda_{1s2p}$ is the {\it incoherence fraction}. $\delta$
is the Dirac-delta function.

The total half width is given by

\begin{equation}
\Gamma_{2p} (\Delta \omega) = \gamma_{2p}(\Delta \omega) +
\Gamma_{2p1s}/2
\end{equation}
where $\Gamma_{2p1s}$ is the radiative decay rate in Lyman $\alpha$
and
$\gamma_{2p}(\Delta \omega)$ is the half width due to collisional
broadening (for clarity, other broadening mechanisms are neglected
here). Expressions for $\gamma_{2p}(\Delta \omega)$ are given by
\mylongcite{Rocket}. The incoherence fraction is given by

\begin{equation}
\Lambda_{1s2p}(\Delta \omega) = \frac{\gamma_{2p}(\Delta
\omega)}{\Gamma_{2p}(\Delta
\omega)} .
\end{equation}

The incoherence fraction is the probability that an atom will undergo
an elastic collision (which randomly changes the energy and thus 
destroys coherence) before it decays. Thus Equation~(2) means that a
fraction $\Lambda$ of scattering processes will undergo complete
redistribution and a fraction $1 - \Lambda$ will undergo no
redistribution.
The incoherence fraction is clearly the physical quantity which
determines the importance of PRD effects. It always lies between 0 and
1. If is it 0 then the transition in question exhibits pure Rayleigh
(coherent) scattering, while if it is 1 the system displays complete
redistribution (CRD). Thus PRD effects are only likely to be important
for values of the incoherence fraction that are $\ll 1$.

The incoherence fraction is determined by the broadening parameter
$\gamma_{2p}(\Delta \omega)$. This quantity is discussed in detail by
\mylongcite{Yelnik} and \mylongcite{CBH}. It is obvious that
$\gamma_{2p}(\Delta \omega)$ should be density dependent, since
a lower density will lead to fewer collisions and hence a smaller
incoherence fraction. Therefore, it is to be expected that PRD should be
significantly more important in a giant star (such as $\beta$~Gem
which has a surface gravity of $5.6 \times 10^{2}$~cm~s$^{-2}$) than
in a main sequence star like the Sun (surface gravity $2.75 \times
10^{4}$~cm~s$^{-2}$). It is for this reason that it is of particular
interest to examine the trend of PRD effects in different types of
star.

The treatment of the Lyman $\beta$ line is further complicated by the possibility of resonant Raman scattering involving the Balmer $\alpha$ line. When a Balmer $\alpha$ photon is absorbed, exciting the $n=3$ level, the $n=3$ level may subsequently decay to $n=1$ emitting a Lyman $\beta$ photon. The frequency of the emitted Lyman $\beta$ photon will be coupled to the frequency of the absorbed photon, unless a collision occurs to redistribute the $n=3$ electron energy. A reliable treatment of the Lyman $\beta$ line must therefore account not only for significant coherence in resonant scattering in the Lyman $\beta$ line, but also for the effects of this so-called {\it cross redistribution} between Balmer $\alpha$ and Lyman $\beta$. Equations (47) to (49) in \mylongcite{HL} show how cross redistribution appears in the emission coefficient of Lyman $\beta$.

An additional problem occurs in the application of the Unified Theory to 
hydrogen -- the $l$-degeneracy of the energy levels means that 
the {\it Isolated Line} approximation is not valid. A more general 
theory  is developed 
by \mylongcite{KB1} and Burnett \& Cooper
(\myfirstcite{KB2},b). This predicts corrections to the 
Unified theory which take the form of additional correlation terms in 
the emission coefficients. These terms represent the collisional 
mixing of $l$-degenerate levels during far wing emission and are 
discussed in the context of stellar atmospheres by \mylongcite{CBH}. 
These terms are known to cancel out if the $l$-degenerate levels are 
populated according to their statistical weights. It is known that in 
solar models the $l$-degenerate levels are populated according to their 
statistical weights in the chromosphere and so these correlations terms 
were not included in the study made by \mylongcite{HL}. These correlation 
terms have also been discarded in the present work, but it is noted that, 
in principle, they may become more significant in the lower gravity stars 
considered here since the collision rates between the $l$-degenerate levels 
will be lower due to the lower chromospheric particle densities. Estimates 
of the departure from population of $l$-degenerate levels according 
to statistical weights have been made and are discussed in Section~4.

\section{Models}
\label{Modsec}

PRD calculations have been performed using atmospheric models of three
different stars, the low gravity giant $\alpha$~Tau (K5~III), the
higher gravity giant $\beta$~Gem (K0~III) and the sub-giant Procyon
(F5~IV-V). Basic parameters for these stars are given in
Table~\ref{Param}. 

\begin{table}
\begin{center}
\caption{Fundamental stellar parameters.}
\begin{tabular}{|c||c|c|c|c|} \hline
Parameter & $\alpha$~Tau & $\beta$~Gem & Procyon \\ \hline  
Catalogue No. & HD 29139 & HD 62509 & HD 61421 \\
Spectral Type & K5
III & K0 III & F5 IV-V \\  
$T_{\mbox{\scriptsize eff}}$ & 3920 K$^{a}$
& 4865 K$^{e}$& 6500 K$^{g}$ \\  
$\log g_{*}$ & 1.25$^{b}$ &
2.75$^{e}$ & 4.00$^{h}$  \\  
Radius & 44.3 $R_{\odot}$$^{c}$ & 8.93
$R_{\odot}$$^{f}$  & 2.1 $R_{\odot}$$^{i}$ \\  
Dist. from Sun & 19.96
pc$^{d}$& 10.3 pc$^{d}$ & 3.5 pc$^{i}$\\ \hline
\label{Param}
\end{tabular}
\end{center}

{ $^{a}$ Blackwell, Lynas-Gray \& Petford (\myfirstcite{BLP91})

$^{b}$ \mylongcite{BnB93}

$^{c}$ Calculated using \mylongcite{BLP91}.

$^{d}$ From Hipparcos (ESA 1997).

$^{e}$ \mylongcite{Drake91}

$^{f}$ Calculated using \mylongcite{Mozurkewich}.

$^{g}$ \mylongcite{Ayres}

$^{h}$ \mylongcite{Drake95}

$^{i}$ \mylongcite{Wood96}}

\end{table}

The $\alpha$~Tau model adopted is that of \mylongcite{Mcmurry99}. This
model was developed from the \mylongcite{Kelch} chromospheric model
and an observed emission measure distribution for the transition
region. It was contructed using an approximate PRD approach, based on an
truncation of the Lyman~$\alpha$ profile. The properties of the high 
temperature material in $\alpha$ Tau are not well constrained 
observationally. The atmospheric model only extends out to a column mass 
density of $10^{-5}$g~cm$^{-2}$ and a temperature $T_{\mbox e} = 10^{5}$K 
which means that the optical depth in both Lyman $\alpha$ and Lyman $\beta$ 
is still significant at the top point in the atmosphere. When using this 
model to predict observations, the atmosphere is extrapolated outwards to 
zero optical depth, assuming a constant source function. This estimate of 
the behaviour of the atmosphere above the top of the model limits the 
reliability of the calculated line core profiles, up to the wavelength 
difference from line centre where the optical depth at the top point in 
the atmosphere becomes negligible, which occurs at a detuning of 
around 0.4\AA~for Lyman $\alpha$ and 0.3\AA~for Lyman $\beta$. This 
extrapolation is significant in the calculated integrated line fluxes, 
particularly the Lyman $\beta$ CRD flux which is dominated by the flux 
in the optically thick line core. The $\alpha$ Tau model does not account 
for the stellar wind which will contain a significant fraction of neutral 
hydrogen. The wind will introduce an additional source of optical depth 
above the top of the atmospheric model which further limits the physical 
reality of the upper boundary conditions in the model. 
Recent {\it Far Ultraviolet Spectroscopic Explorer} ({\it FUSE}) observations of $\alpha$~Tau will help to constrain the properties 
of the upper atmosphere of the star.

\begin{figure}
\begin{center}
\centerline{\vbox{\psfig{file=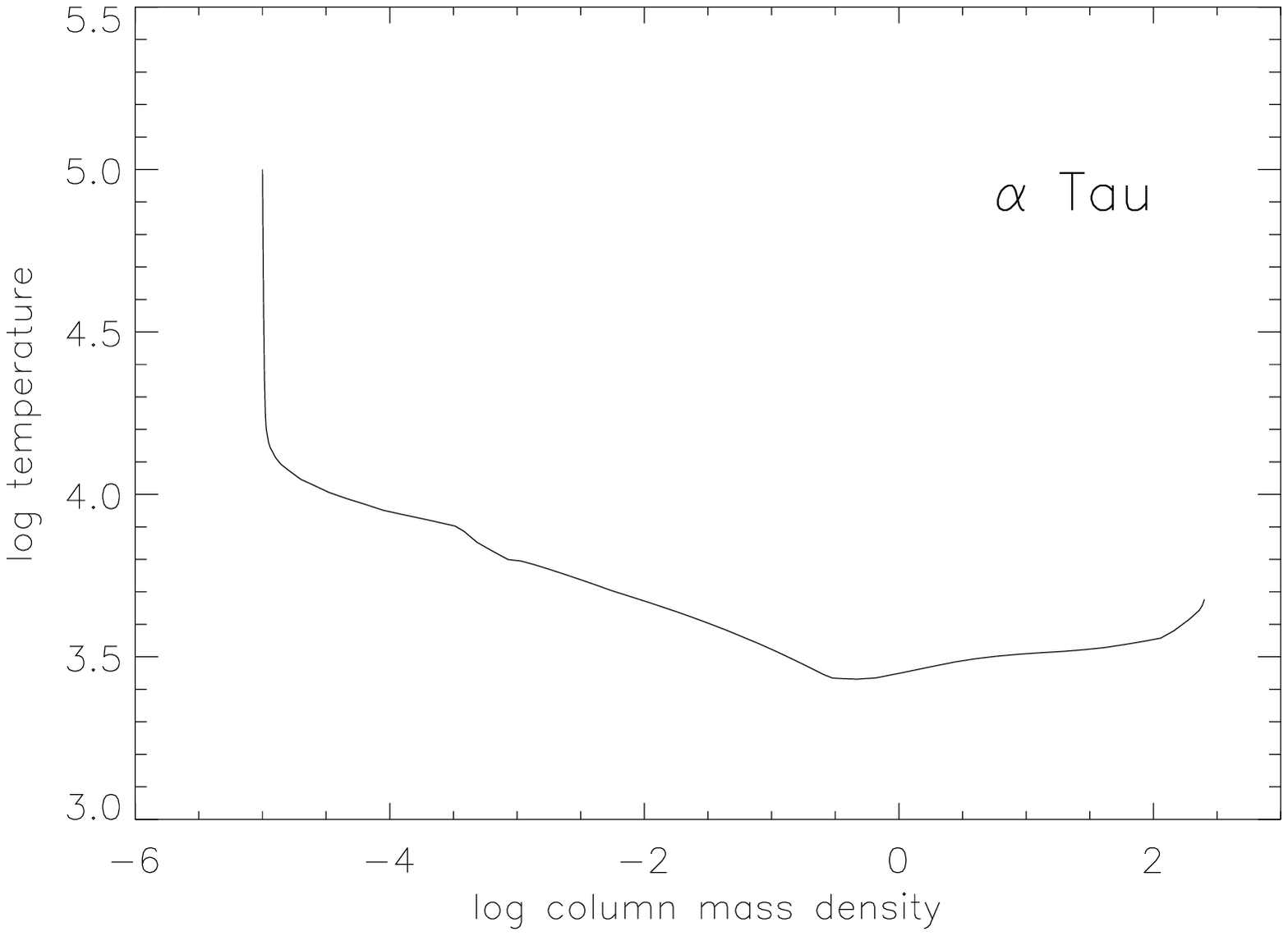,width=8.0cm,height=7.0cm}
\psfig{file=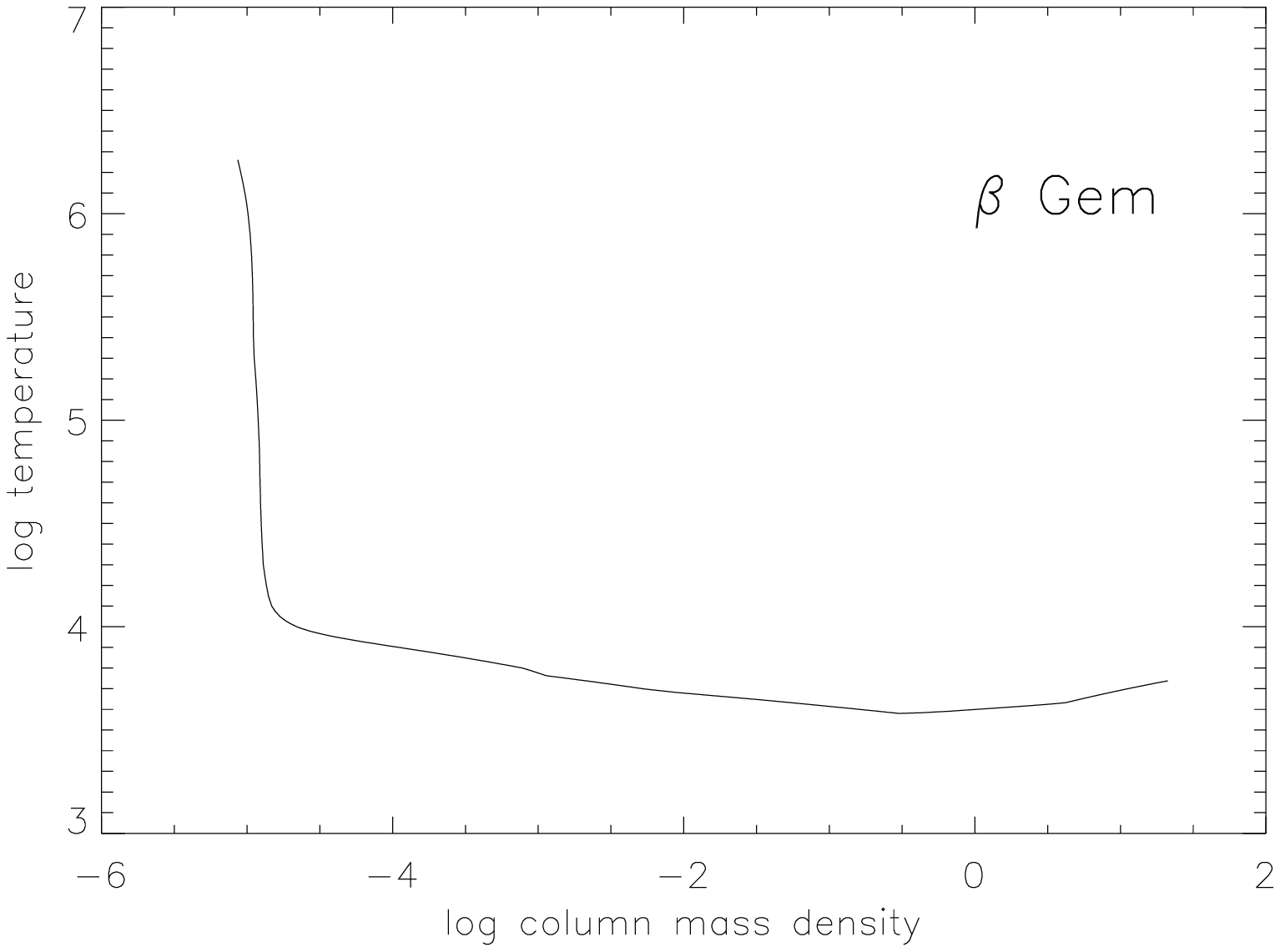,height=7.0cm,width=8.0cm}
\psfig{file=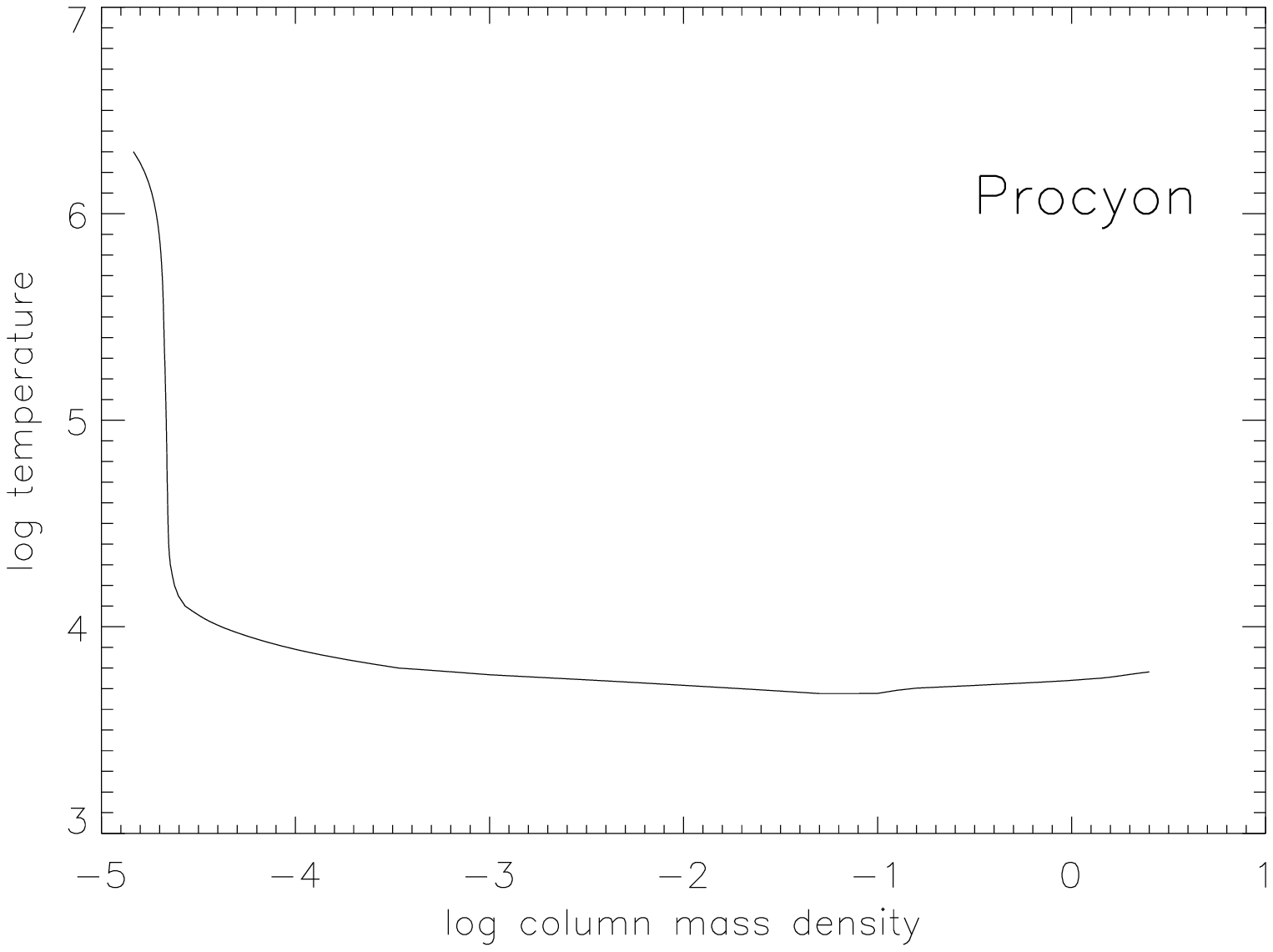,height=7.0cm,width=8.0cm}}}
\caption{The atmospheric models. In each case the electron temperature
is given in K and the column mass density is given in g~cm$^{-2}$.}
\label{atmos}
\end{center}
\end{figure}

The $\beta$~Gem
and Procyon models are new and have been constructed by the author in
a similar fashion to the $\alpha$~Tau model. The models are based on
the chromospheric models of \mylongcite{Kelch} and Ayres, Linsky \&
Shine (\myfirstcite{Ayres}) respectively, and observed emission measure
distributions. These new models are initially
constructed using the assumption of CRD. The details of the
construction of these models will be discussed in future papers. The 
new models extend to coronae with $\log T_{\mbox{\scriptsize e}} = 6.3$. The Lyman lines 
are optically thin at the top of these coronal models and so the 
accuracy of the calculated line cores are not limited by an 
extrapolation in the same way as in $\alpha$~Tau. The
electron temperature as a function of the column mass density is shown
for each of the models in Figure~\ref{atmos}.

\section{Results and Discussion}

The Hubeny~\&~Lites version of the MULTI code has been applied to each
of the three atmospheric models described in Section~\ref{Modsec}.
Each model has been run for four different cases, the same cases that
were studied by \mylongcite{HL}, namely: 

\begin{description}
\item{(a)} CRD
\item{(b)} PRD resonance scattering in Lyman~$\alpha$,
\item{(c)} PRD resonance scattering in Lyman~$\alpha$ and
Lyman~$\beta$,
\item{(d)} PRD resonance scattering in Lyman~$\alpha$ and
Lyman~$\beta$, and cross-redistribution between Balmer~$\alpha$ and 
Lyman~$\beta$.
\end{description}

For all twelve of these calculations, the physical depth and frequency
dependent coherence fractions were used. Balmer~$\alpha$ is treated in CRD 
throughout
(\mycite{HnH84} show that PRD is not significant in determining the 
line profile for a subordinate line, provided that the damping parameters 
for the upper and lower levels are not too different).  The calculations 
were performed using a nine-level
atomic model of hydrogen (the same hydrogen model which was used by
\mycite{Mcmurry99}). This model includes 28 bound-bound transitions and
8 free-bound transitions. The Lyman continuum is calculated in detail
while the others are treated using radiation temperatures. There is no
fine structure in the atomic model (the fine structure splitting
between components is much smaller than the width of the Lyman lines
and is not observable).

As discussed in Section~2, the PRD treatment employed here neglects
certain correlation terms, the importance of which depends upon how
closely the $l$-degenerate levels are populated in accordance with
their statistical weights (if the $l$-degenerate levels are populated
exactly according to their statistical weights these correlation terms
have no effect). In order to attempt to estimate the relative
populations of the $l$-degenerate levels basic statistical equilibrium
calculations have been applied to a simplified fine structure
model of hydrogen. This model consists of only six levels (1s, 2s, 2p,
3s, 3p and 3d). The transition rates between these levels due to
spontaneous emission, stimulated emission, radiative absorption and
electron collisions (both elastic and inelastic) were calculated for
an electron temperature $T_{e} = 10^{4}$K (and the appropriate
electron density for each of the three stellar models) using the
radiation field from MULTI calculations, the estimates for the elastic
collision rates given by \mylongcite{CBH} and the inelastic collision
rates calculated by \mylongcite{GnP} (note that while Chang, Avrett \&
Loeser~\myfirstcite{CAL} show that there are inconsistencies in the
rates calculated by \mycite{GnP} involving states with $n \geq 4$
they do not find significant problems with the fine structure rates
between states with $n \leq 3$). Under the conditions used for each of
the three stars it was found that spontaneous emission, radiative
absorption and elastic collisions are the important processes while
stimulated emission and inelastic collisions are not important under
these conditions. The calculated relative level populations for each
of the fine structure levels in $n=2$ and $n=3$ are given in
Table~\ref{finestruc}. The ratios of statistical weights for the fine
structure levels are $\omega_{\mbox{\scriptsize 2p}} /
\omega_{\mbox{\scriptsize 2s}} = 3$, $\omega_{\mbox{\scriptsize
3p}} / \omega_{\mbox{\scriptsize 3s}} = 3$ and
$\omega_{\mbox{\scriptsize 3d}} / \omega_{\mbox{\scriptsize 3s}} =
5$. The results shown in Table~\ref{finestruc} suggest that the
fine structure level population ratios are likely to be very close to
the ratios of the statistical weights (the greatest variation being 4
per cent in the ratio of the 3d to 3s populations) and so the
correlation terms may be safely neglected under conditions typical of
the region where PRD effects are important in these stars. When
considering these results, it must be borne in mind that the fine
structure model that has been used is very simple: transitions between
levels with principle quantum number greater than $n=3$ have been
ignored. It is estimated that, for example, radiative decays from
$n=4$ to $n=3$ will populate the $n=3$ at a rate of around ten per
cent of the rate at which $n=3$ is populated by radiative absorption
from $n=2$ (which is the dominant excitation mechanism for $n=3$ under
the conditions in the simple fine structure model).

\begin{table}
\begin{center}

\caption{Relative level populations for fine structure levels. The 
notation $n_{\mbox{\scriptsize $nl$}}$ denotes the population of 
fine structure level $nl$.}

\begin{tabular}{cccc} \\ \hline
Star 
& $n_{\mbox{\scriptsize 2p}}/n_{\mbox{\scriptsize 2s}}$
%& $\frac{(n_{\mbox{\scriptsize 2p}} - 3 n_{\mbox{\scriptsize 2s}})}{n_{2}}$ 
& $n_{\mbox{\scriptsize 3p}}/n_{\mbox{\scriptsize 3s}}$ 
%& $\frac{(n_{\mbox{\scriptsize 3p}} - 3 n_{\mbox{\scriptsize 3s}})}{n_{3}}$ 
& $n_{\mbox{\scriptsize 3d}}/n_{\mbox{\scriptsize 3s}}$ \\ \hline
%& $\frac{(n_{\mbox{\scriptsize 3d}} - 5 n_{\mbox{\scriptsize 3s}})}{n_{3}}$ \\ \hline

$\alpha$ Tau & 2.97 & 3.09 & 4.95 \\
$\beta$ Gem & 2.99 & 3.03 & 4.83 \\
Procyon & 3.00 & 3.00 & 4.94 \\ \hline
\label{finestruc}

\end{tabular}
\end{center}
\end{table}

\begin{figure}
\begin{center}
\centerline{\vbox{
\psfig{file=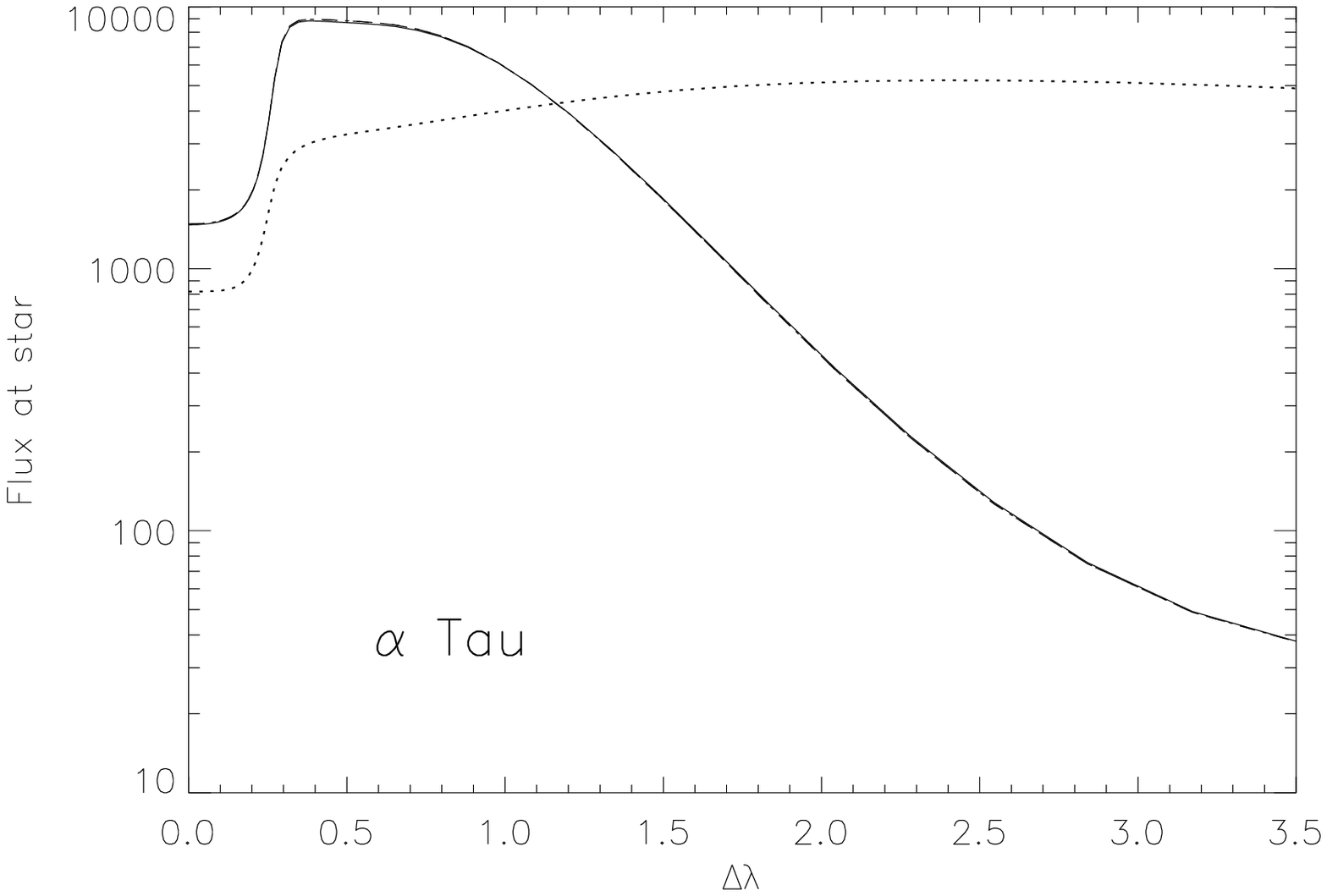,height=7.0cm,width=8.0cm}
\psfig{file=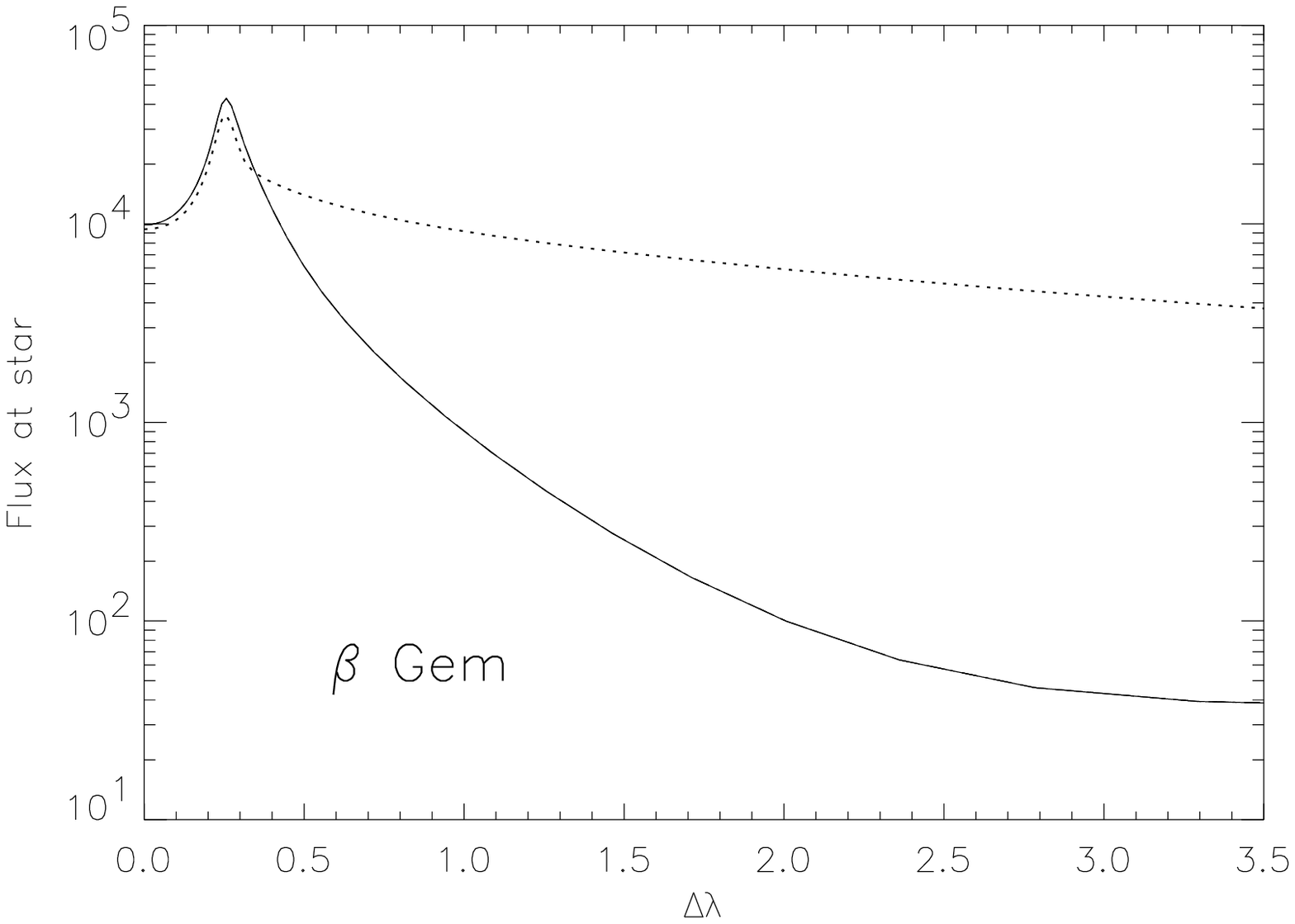,height=7.0cm,width=8.0cm}
\psfig{file=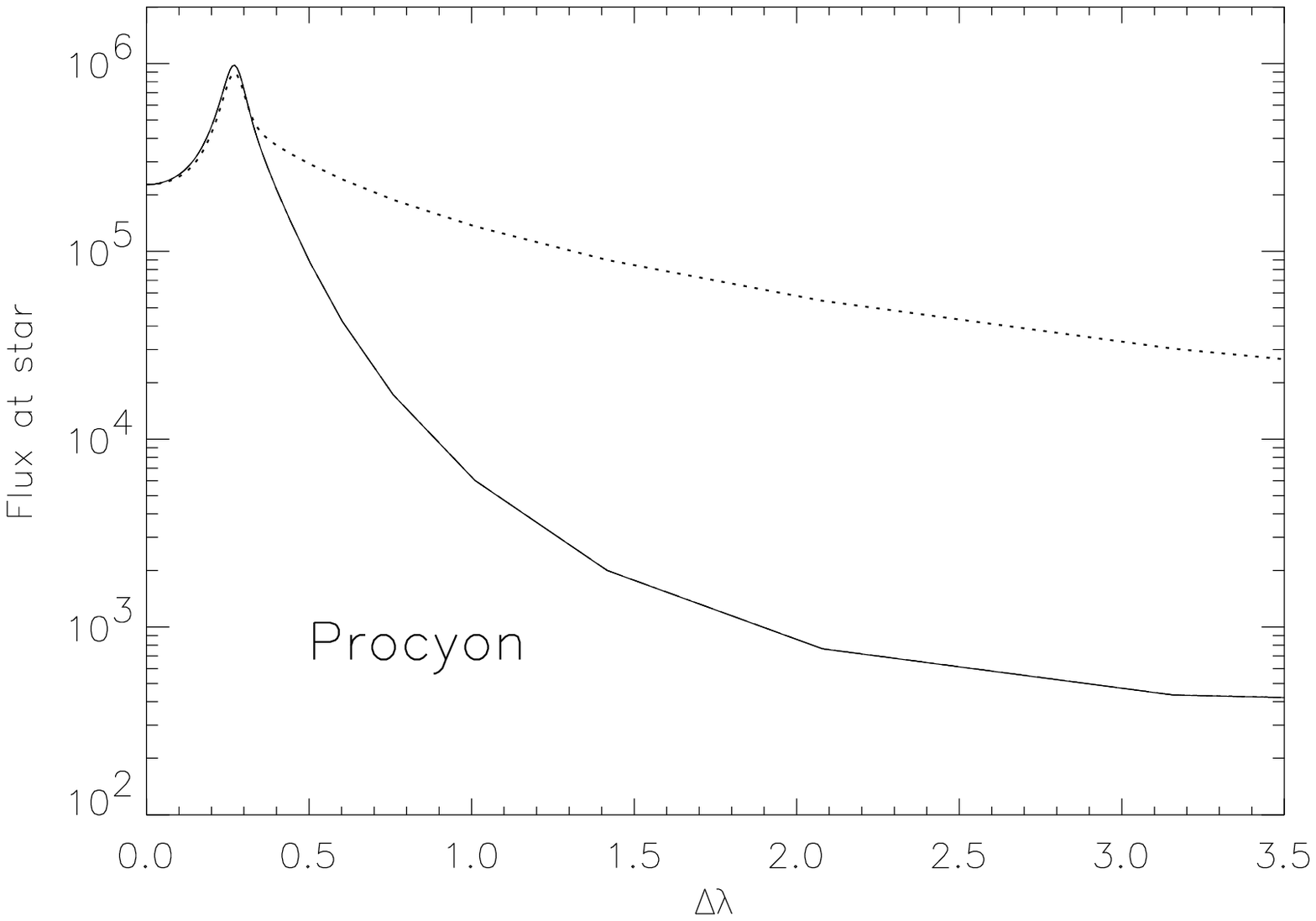,height=7.0cm,width=8.0cm}}}
\caption{Calculated Lyman~$\alpha$ profiles. The dotted line is case
(a), the broken line is case (b), the dot-dash line is case (c) and
the solid line is case (d) (some cases are indistinguishable: see
text). All fluxes are in ergs~cm$^{-2}$s$^{-1}$\AA$^{-1}$ at the stellar
surface. $\Delta \lambda$ is the wavelength relative to line centre in
\AA.}
\label{acal}
\end{center}
\end{figure}

\subsection{Line Profiles}

\begin{figure}
\begin{center}
\centerline{\vbox{
\psfig{file=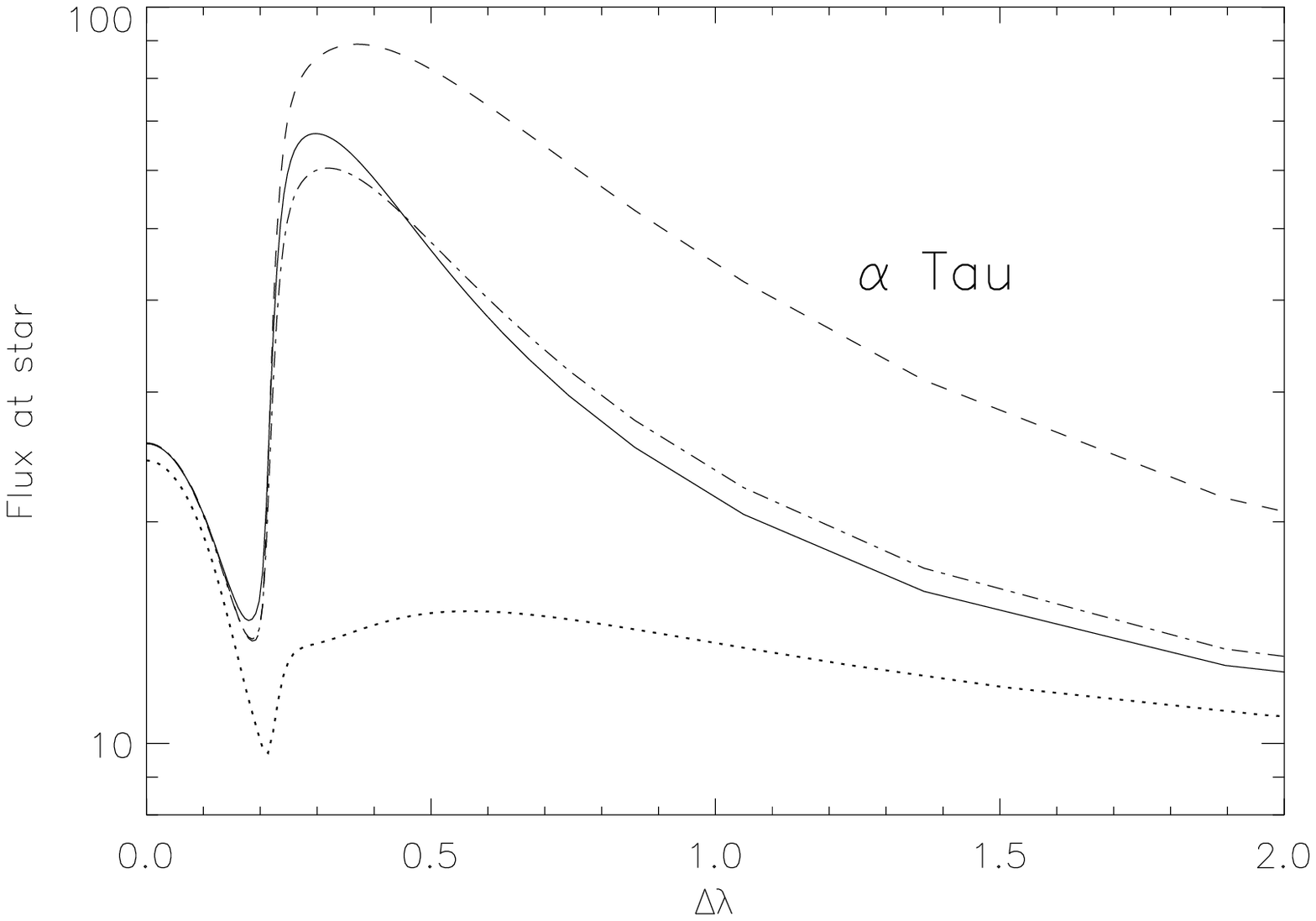,height=7.0cm,width=8.0cm}
\psfig{file=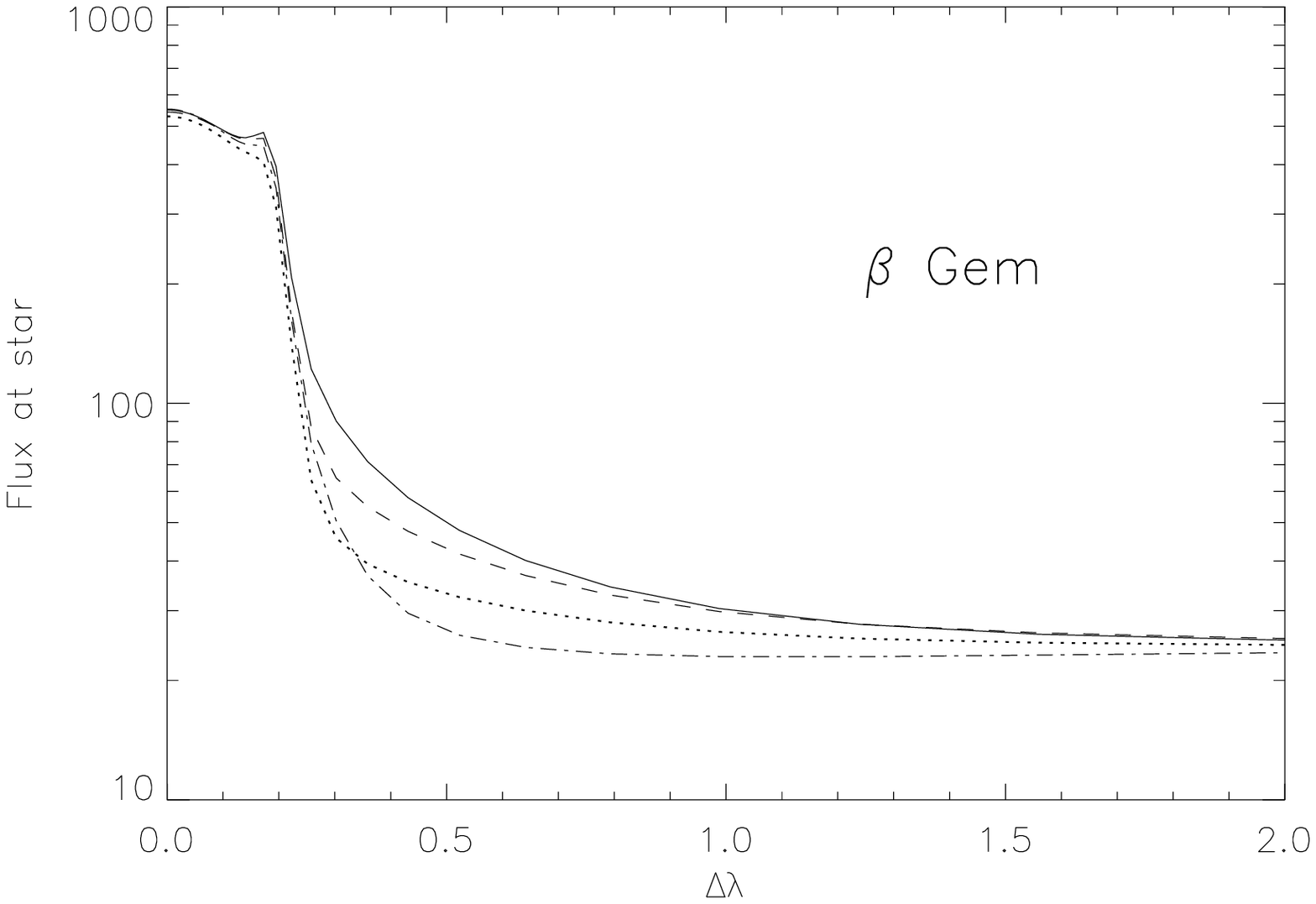,height=7.0cm,width=8.0cm}
\psfig{file=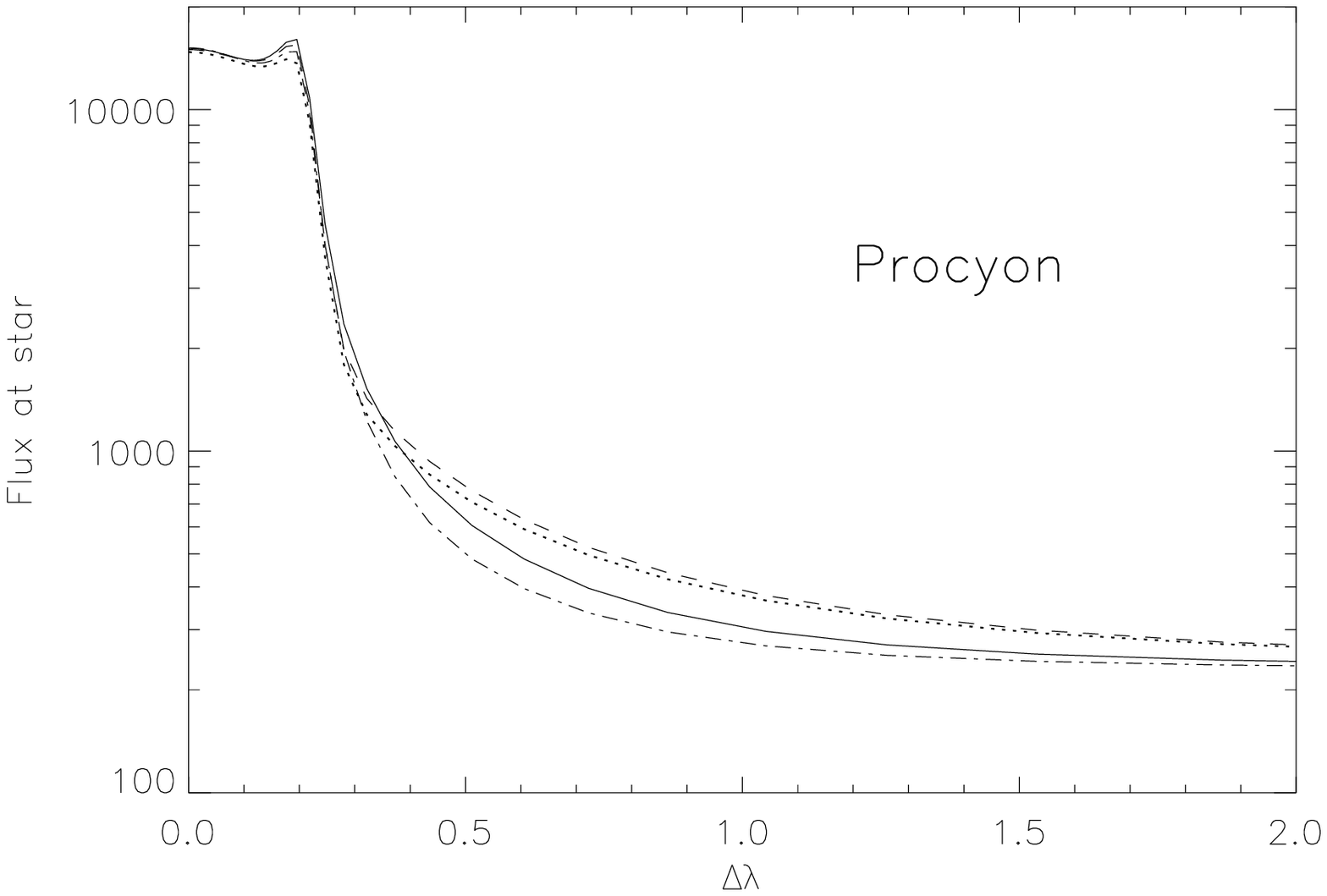,height=7.0cm,width=8.0cm}}}
\caption{Calculated Lyman~$\beta$ profiles. The dotted line is case
(a), the broken line is case (b), the dot-dash line is case (c) and
the solid line is case (d). All fluxes are  in
ergs~cm$^{-2}$s$^{-1}$\AA$^{-1}$ at the stellar surface. $\Delta
\lambda$ is the wavelength relative to line centre in \AA.}
\label{bcal}
\end{center}
\end{figure}

The calculated Lyman~$\alpha$ profiles for the four cases of each of
the models are shown in Figure~\ref{acal} and the corresponding
Lyman~$\beta$ profiles are shown in Figure~\ref{bcal}. These plots
confirm that the treatment of redistribution has a significant effect
upon the calculated hydrogen line profiles. Most of the differences 
between PRD and CRD results is due to the effect of coherent 
scattering (in the atom frame) in Lyman $\alpha$ which is included 
by PRD but not by CRD.  It is clear that the
primary effect of a PRD treatment rather than a CRD treatment is a
significant reduction in the strength of the Lyman $\alpha$ line
wings. This effect occurs because the low incoherence fraction in PRD 
greatly reduces the number of photons that are scattered into the line 
wings compared to complete incoherence in a CRD treatment. In all three 
stars considered, the
treatment of Lyman~$\beta$ does not have a significant effect on the
Lyman~$\alpha$ profile (for each star cases (b), (c) and (d) give
identical Lyman~$\alpha$ profiles), however the treatment of
Lyman~$\alpha$ does strongly effect the calculated Lyman~$\beta$
profile. Coherent scattering (in the atom frame) in Lyman~$\alpha$ affects the
calculated Lyman~$\beta$ as strongly as coherent scattering in
Lyman~$\beta$ -- coherent scattering in Lyman~$\alpha$ traps photons in 
the Lyman $\alpha$ line which pumps the $n=2$ level population above its 
CRD value, this increases the
$n=3$ level population since radiative excitation from $n=2$ 
to $n=3$ (absorption of Balmer $\alpha$ photons) is a significant 
population process for $n=3$. This increase in the $n=3$ population 
leads to an
increased flux in the Lyman~$\beta$ line wings. This effect is evident in
all the case~(a) Lyman~$\beta$ profiles. The effect of including PRD 
effects in the treatment of
resonant scattering in Lyman~$\beta$ is to reduce the near wing flux
by a similar mechanism to that which operates in Lyman~$\alpha$ (see
above). Finally, the inclusion of cross-redistribution increases the
line wing flux again. These effects were all noted and discussed by
\mylongcite{HL} for the solar case. 

The most striking result from Figures~\ref{acal} and~\ref{bcal} is the
significance of the trend with surface gravity. It is clear that the
importance of PRD effects grows rapidly as stars of lower surface
gravity (and hence density) are examined. This is to be expected
because the elastic collision rates (and hence the incoherence
fraction) are smaller in low density stars. As noted in
Section~\ref{theory}, PRD is most different from CRD when the
incoherence fraction is small, so it is to be expected that the
greatest difference occurs in low surface gravity stars. $\beta$~Gem
and, in particular, $\alpha$~Tau show very much stronger PRD effects
that Procyon or the Sun \myshortcite{HL}. In both cases the  reduction
in the width of the Lyman~$\alpha$ line is extremely significant and
in $\alpha$~Tau the enhancement of the near wings of Lyman~$\beta$ is
sufficient to make them stronger than the line core. Notice
that even in these extreme examples of the importance of PRD, its
inclusion in the treatment of Lyman~$\beta$ is not as
significant as in the treatment of Lyman~$\alpha$. However, the effects 
on Lyman~$\beta$ are still significant and should not be disregarded.

\begin{figure}
\begin{center}
\centerline{\vbox{
\psfig{file=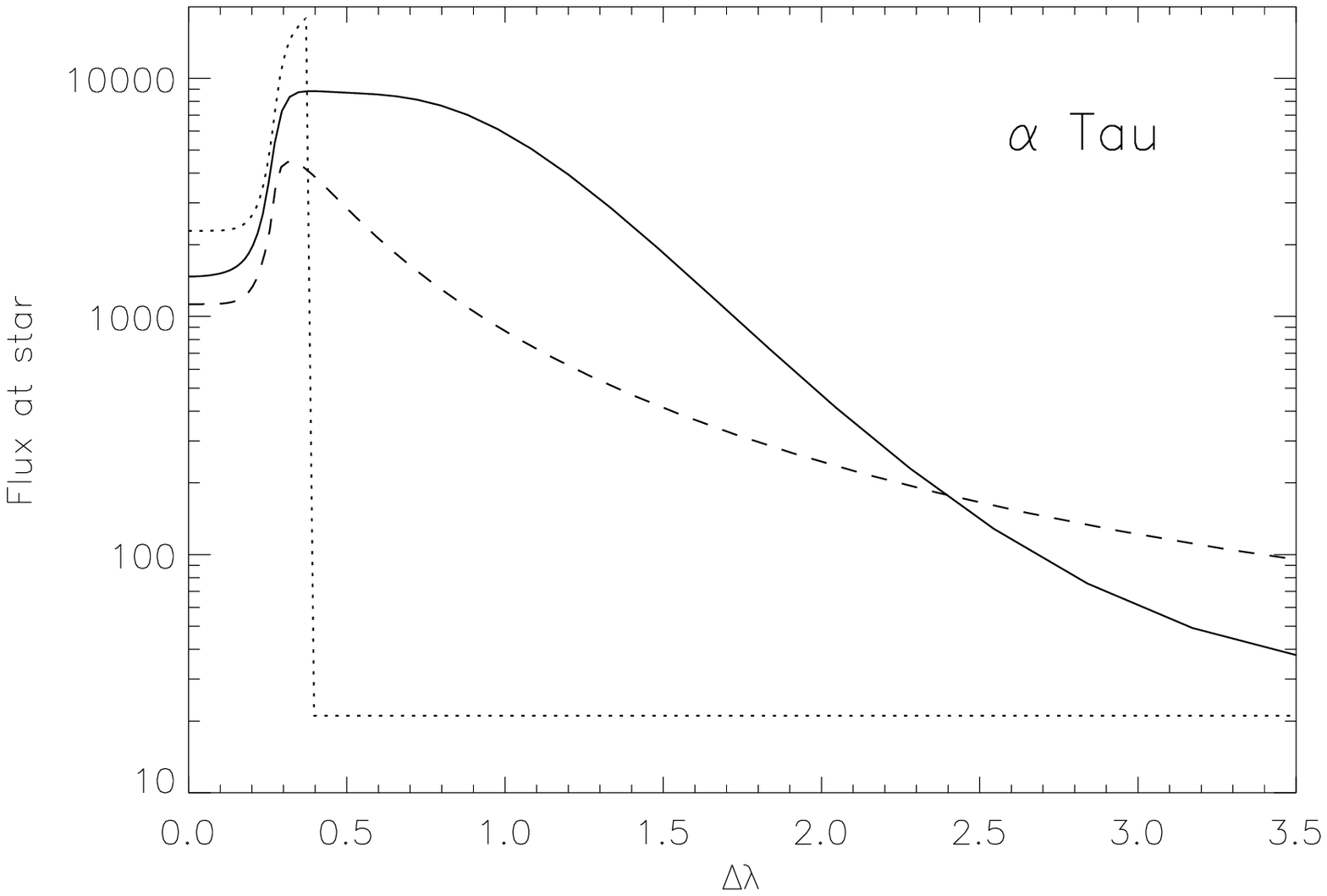,height=7.0cm,width=8.0cm}
\psfig{file=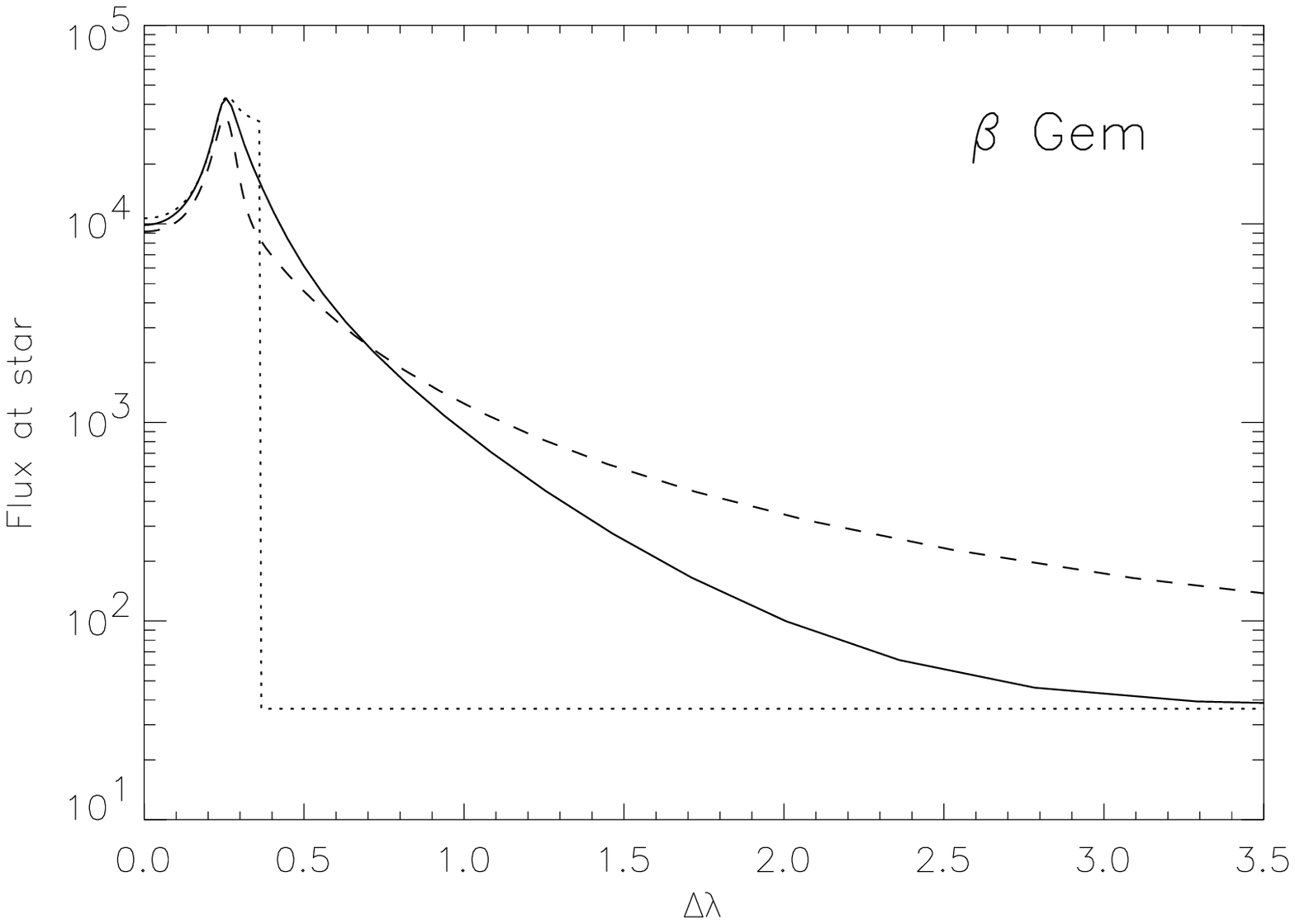,height=7.0cm,width=8.0cm}
\psfig{file=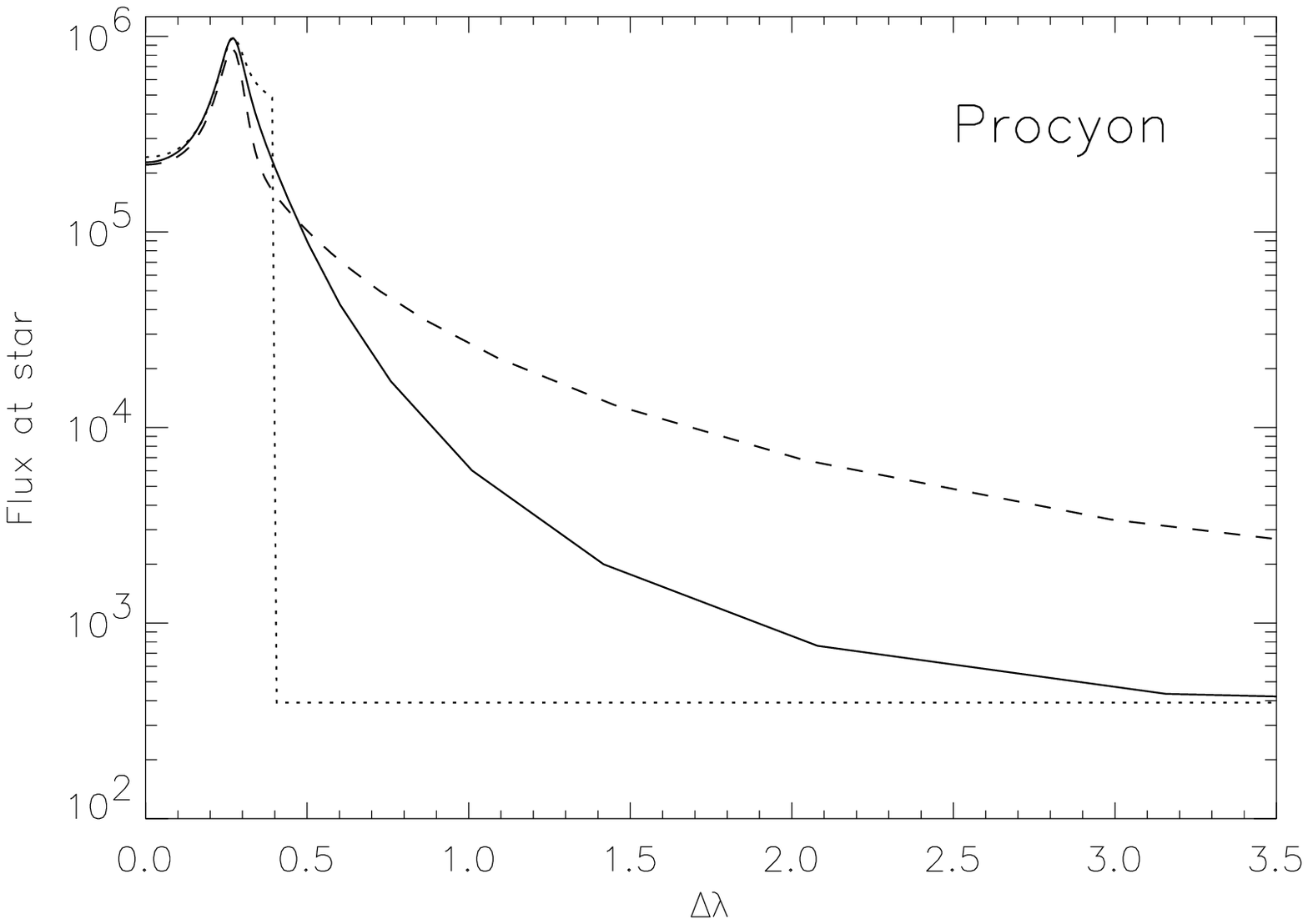,height=7.0cm,width=8.0cm}}}
\caption{Calculated Lyman~$\alpha$ profiles. The solid line is case
(d), the dotted line is case (e) and the dashed line is case
(f). All fluxes are in ergs~cm$^{-2}$s$^{-1}$\AA$^{-1}$ at the stellar
surface. $\Delta \lambda$ is the wavelength relative to line centre in
\AA.}
\label{aapprox}
\end{center}
\end{figure}

\begin{figure}
\begin{center}
\centerline{\vbox{
\psfig{file=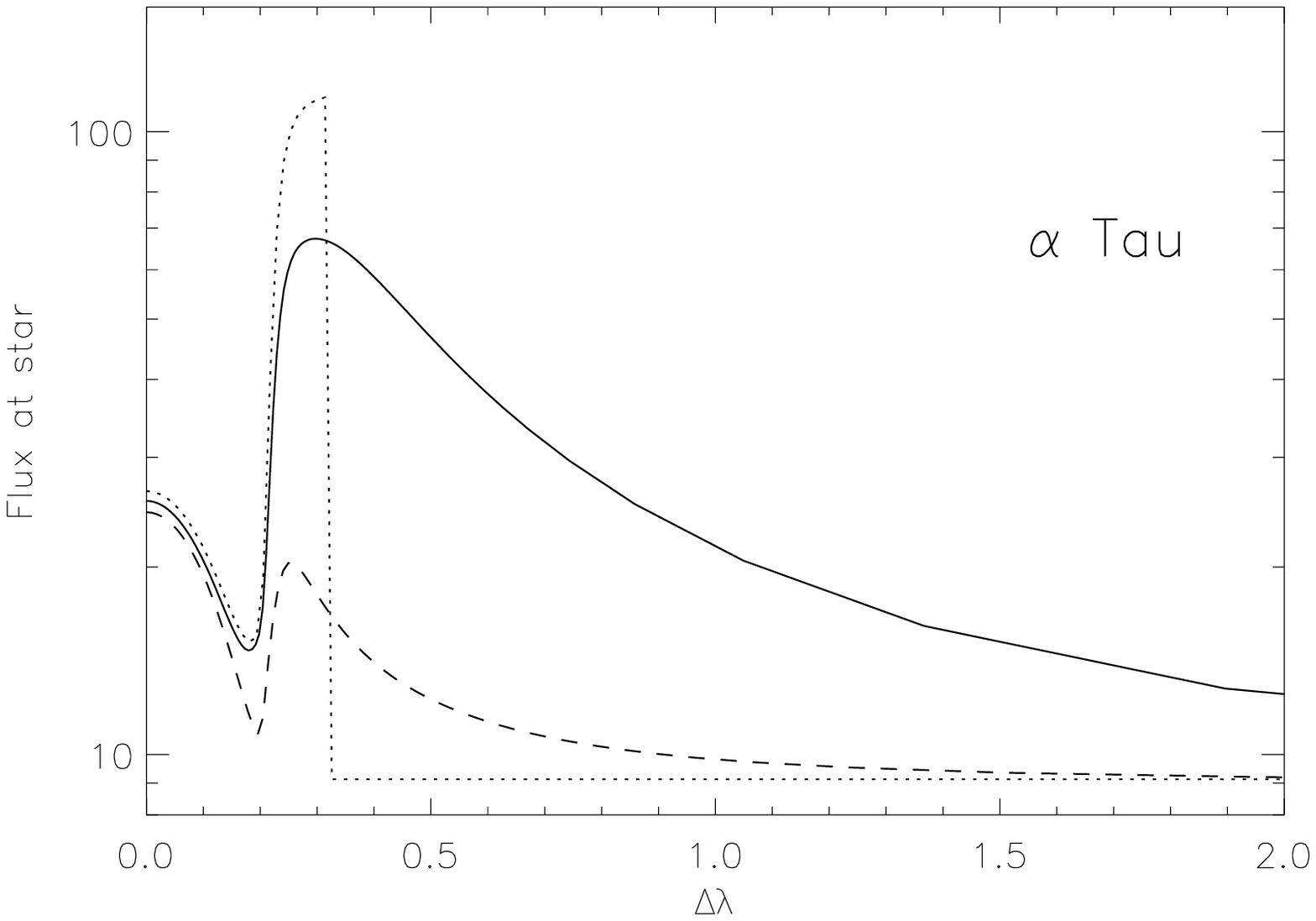,height=7.0cm,width=8.0cm}
\psfig{file=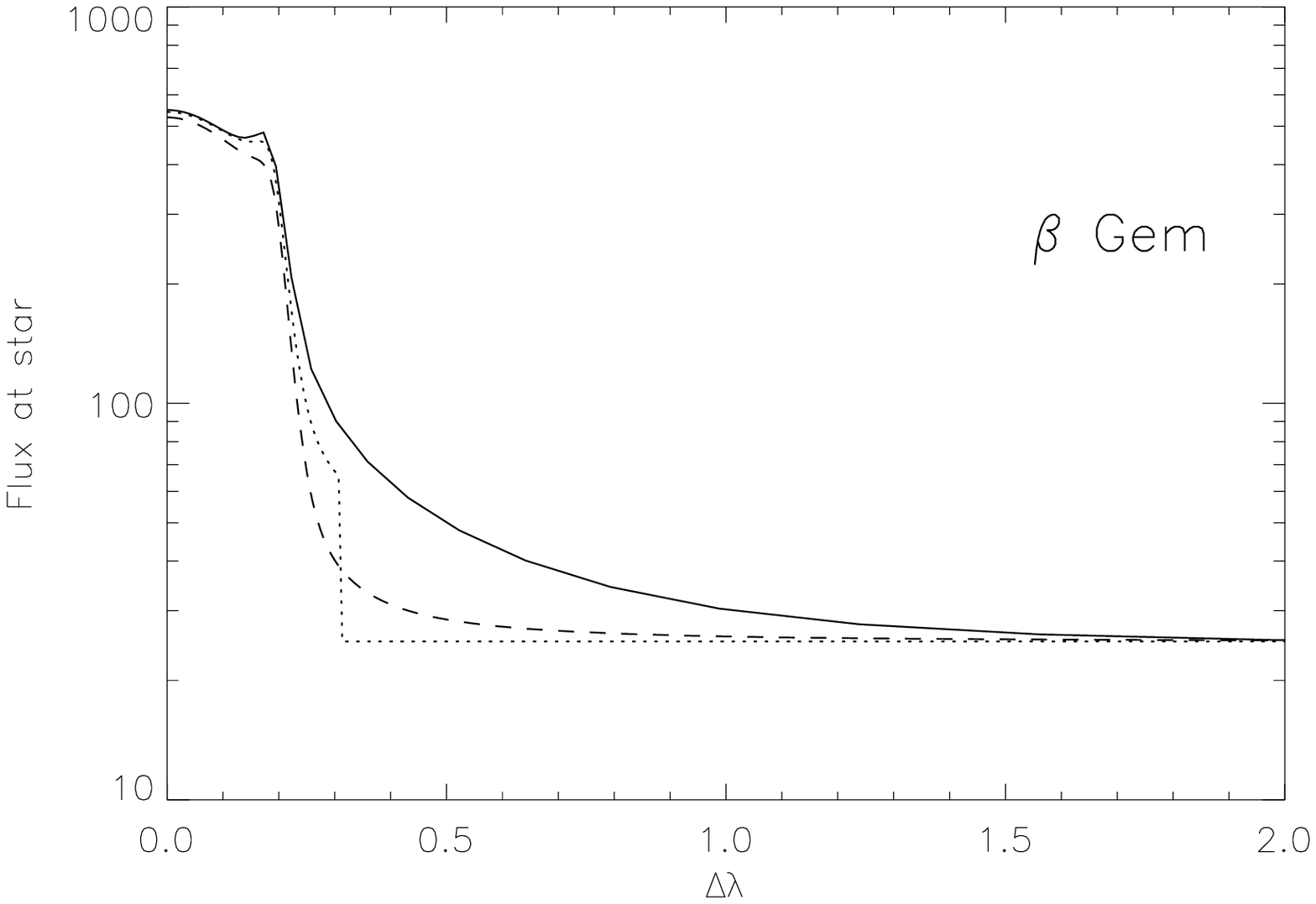,height=7.0cm,width=8.0cm}
\psfig{file=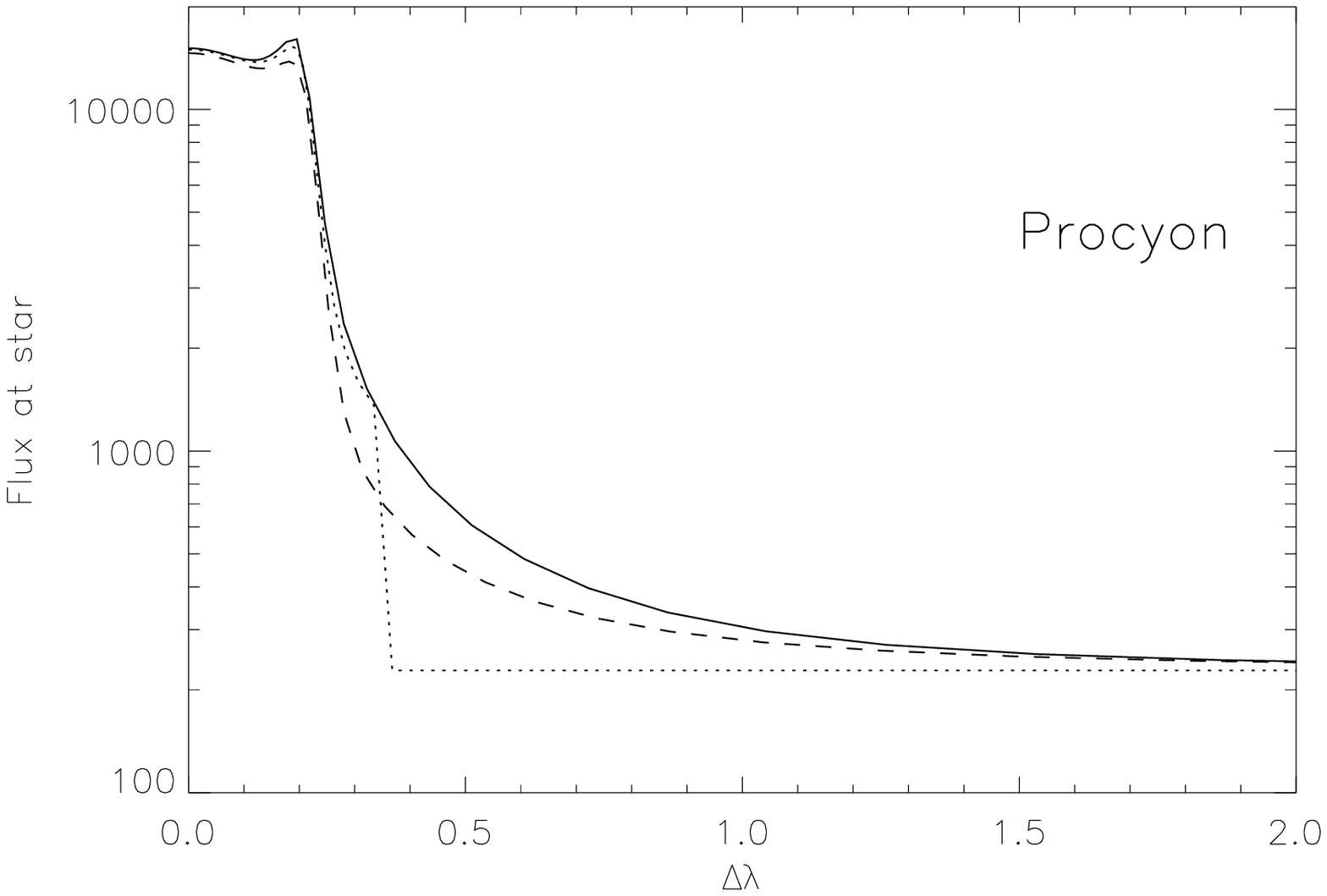,height=7.0cm,width=8.0cm}}}
\caption{Calculated Lyman~$\beta$ profiles. The solid line is case
(d), the dotted line is case (e) and the dashed line is case
(f). All fluxes are in ergs~cm$^{-2}$s$^{-1}$\AA$^{-1}$ at the stellar
surface. $\Delta \lambda$ is the wavelength relative to line centre in
\AA.}
\label{bapprox}
\end{center}
\end{figure}

For comparison the models have also been run in MULTI using two
approximate PRD techniques. The first approximate technique (which
will be referred to as case (e)) is the commonly used method of
simulating PRD effects by performing CRD calculations using a
truncated Voigt absorption profile. In the calculations presented here
the absorption profiles for both Lyman~$\alpha$ and Lyman~$\beta$ have
been truncated at $\Delta \lambda = 6 \Delta \lambda_{D}$ where
$\Delta \lambda_{D}$ is the Doppler width (including thermal and 
microturbulent contributions) at $T_{\mbox{e}} = 8 \times 10^{3}$K. 
This form of truncated profile was used by \mylongcite{GMHATrA} in 
the modelling of the `hybrid' giant $\alpha$~TrA and variations on this
approximate approach to PRD have been used by many authors (including
\mycite{Mcmurry99} in the modelling of $\alpha$~Tau). The second
approximate technique is that suggested by \mylongcite{Gayley}, which
will be referred to as case (f). In that paper, the author suggests
that if an approximate PRD method that only requires CRD calculations
is required, rather than simply truncating the Voigt profile, one
should use a Voigt profile with a depth dependent Voigt parameter,
$a_{\mbox{\scriptsize eff}} = 1.8 / \tau$ where $\tau$ is the mean
optical depth in the line. This approximation is intended to mimic the 
escape probability for first resonance lines, such as Lyman $\alpha$, and 
is not strictly applicable to Lyman $\beta$ (because of the influence 
of Balmer $\alpha$), but it has been applied to both lines for comparison. 
The profiles calculated using these
approximate PRD methods are compared with the case (d) PRD profiles
for Lyman~$\alpha$ and Lyman~$\beta$ in Figures~\ref{aapprox}
and~\ref{bapprox}. 

As noted by \mylongcite{Gayley}, neither of these approximate methods is
expected to make accurate predictions for the line profiles and this
is evident in Figures~\ref{aapprox} and~\ref{bapprox}. For all six
calculated line profiles both the approximate techniques fail to
accurately mimic the PRD profile. Owing to the truncation of the absorption 
profile, case (e) always significantly underpredicts the flux beyond the 
truncation frequency. For Lyman $\alpha$, case (f) underpredicts the 
strength of the inner line wings and overestimates the strength of the 
outer line wings, producing a profile that is broader than the case (d) 
profile. 

\begin{table}
\begin{center}
\caption{Calculated integrated line fluxes from the models. All fluxes 
are at the
stellar surface in ergs~cm$^{-2}$~s$^{-1}$.}
\begin{tabular}{|c|c|c|c|c|c|} \hline
Line & Case &$\alpha$~Tau & $\beta$~Gem & Procyon \\ \hline
Lyman~$\alpha$ & (a) & $8.35 \times 10^{4}$ & $7.26 \times 10^{4}$ &
$9.67 \times 10^{5}$ \\  & (b) & $1.78 \times 10^{4}$ & $2.06 \times
10^{4}$ & $4.06 \times 10^{5}$ \\  & (c) & $1.78 \times 10^{4}$ &
$2.06 \times 10^{4}$ & $4.06 \times 10^{5}$ \\  & (d) & $1.77 \times
10^{4}$ & $2.06 \times 10^{4}$ & $4.07 \times 10^{5}$  \\ & (e) &
$4.70 \times 10^{3}$ & $1.69 \times 10^{4}$ & $3.86 \times 10^{5}$  \\
& (f) & $5.67 \times 10^{3}$ & $1.72 \times 10^{4}$ & $4.19 \times
10^{5}$  \\ \hline

Lyman~$\beta$  & (a) & $2.15 \times 10^{1}$ & $1.99 \times 10^{2}$ &
$7.24 \times 10^{3}$ \\ & (b) & $1.54 \times 10^{2}$ & $2.31 \times
10^{2}$ & $7.68 \times 10^{3}$ \\  & (c) & $7.47 \times 10^{1}$ &
$2.00 \times 10^{2}$ & $7.12 \times 10^{3}$ \\   & (d) & $7.19 \times
10^{1}$ & $2.48 \times 10^{2}$ & $7.56 \times 10^{3}$ \\   & (e) &
$2.36 \times 10^{1}$ & $2.06 \times 10^{2}$ & $6.86 \times 10^{3}$ \\
& (f) & $9.67 \times 10^{0}$ & $1.92 \times 10^{2}$ & $6.65 \times
10^{3}$ \\ \hline   
\label{fluxcal}
\end{tabular}
\end{center}
\end{table}

\subsection{Integrated Line Fluxes}

Table~\ref{fluxcal} shows the calculated integrated line fluxes at the
stellar surface for each of the three models in all six cases considered. 
These fluxes show that
not only the line profile but also the total flux is significantly
affected by a PRD treatment; the total flux shows differences of up to
a factor of 4.7 between CRD and PRD calculations. 

The main application of the approximate PRD techniques is to calculate 
total line fluxes (as discussed above it is known that they do not 
predict PRD line profiles correctly). Comparing the PRD
integrated line fluxes (case (d)) to those calculated using the 
approximate methods (cases
(e) and (f)) shows that both approximate methods consistently
underpredict the integrated line flux (the only exception being 
Lyman $\alpha$ in Procyon case (f)). The approximate methods work best
in Procyon (where the line fluxes agree with the PRD calculations to 
within 15 per cent) but the discrepancy is larger in the lower
gravity stars: in $\alpha$~Tau, case (e) underpredicts the
Lyman~$\alpha$ flux by a factor of 3.8 and the Lyman $\beta$ flux by a factor 
of 3.
 It is not surprising that these
approximate methods, which are based on CRD, work best in Procyon where
PRD effects are least important. The fluxes given in Table~\ref{fluxcal}
indicate that for estimating the Lyman $\alpha$ integrated line flux, 
case (f) is better than case (e), while for estimating the integrated 
Lyman $\beta$ flux, case (e) is better (this is not surprising since, 
as discussed above, case (f) is not appropriate for Lyman $\beta$). It 
is concluded that these
approximate techniques can be useful in simulating PRD effects in the 
higher gravity stars:
for Procyon, the case (f) Lyman~$\alpha$ flux agrees with
case (d) to within 3 per cent and for $\beta$~Gem, the case (f) 
Lyman $\alpha$ flux agrees with
case (d) to within 20 per cent, compared to a discrepancy between the
flux predicted by normal CRD and case (d) PRD of a factor of 2 in
Procyon and a factor of 3.5 in $\beta$~Gem. The approximate techniques
appear to be less reliable in the lower gravity star $\alpha$ Tau where 
case (f) underpredicts the Lyman $\alpha$ flux by a factor of 3.12 and 
case (e) underpredicts the Lyman $\beta$ flux by a factor of 3.05. However, 
as mentioned in Section~3, the significance of the integrated line fluxes 
in the $\alpha$ Tau model is limited by the reliability of the estimated 
line core profile produced by the extrapolation of the source function 
above the top point in the model atmosphere.

\begin{figure}
\begin{center}
\centerline{\vbox{
\psfig{file=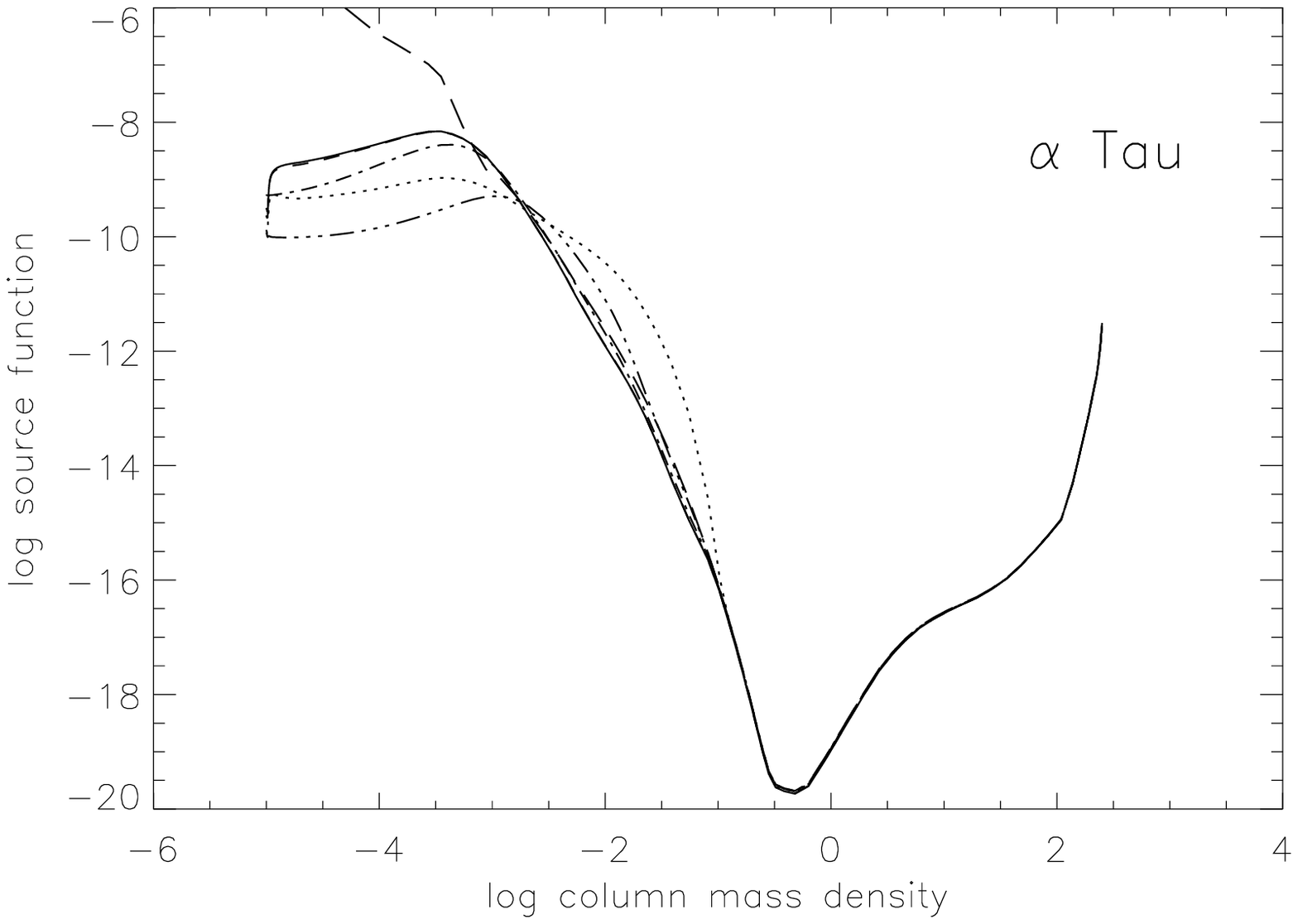,height=7.0cm,width=8.0cm}
\psfig{file=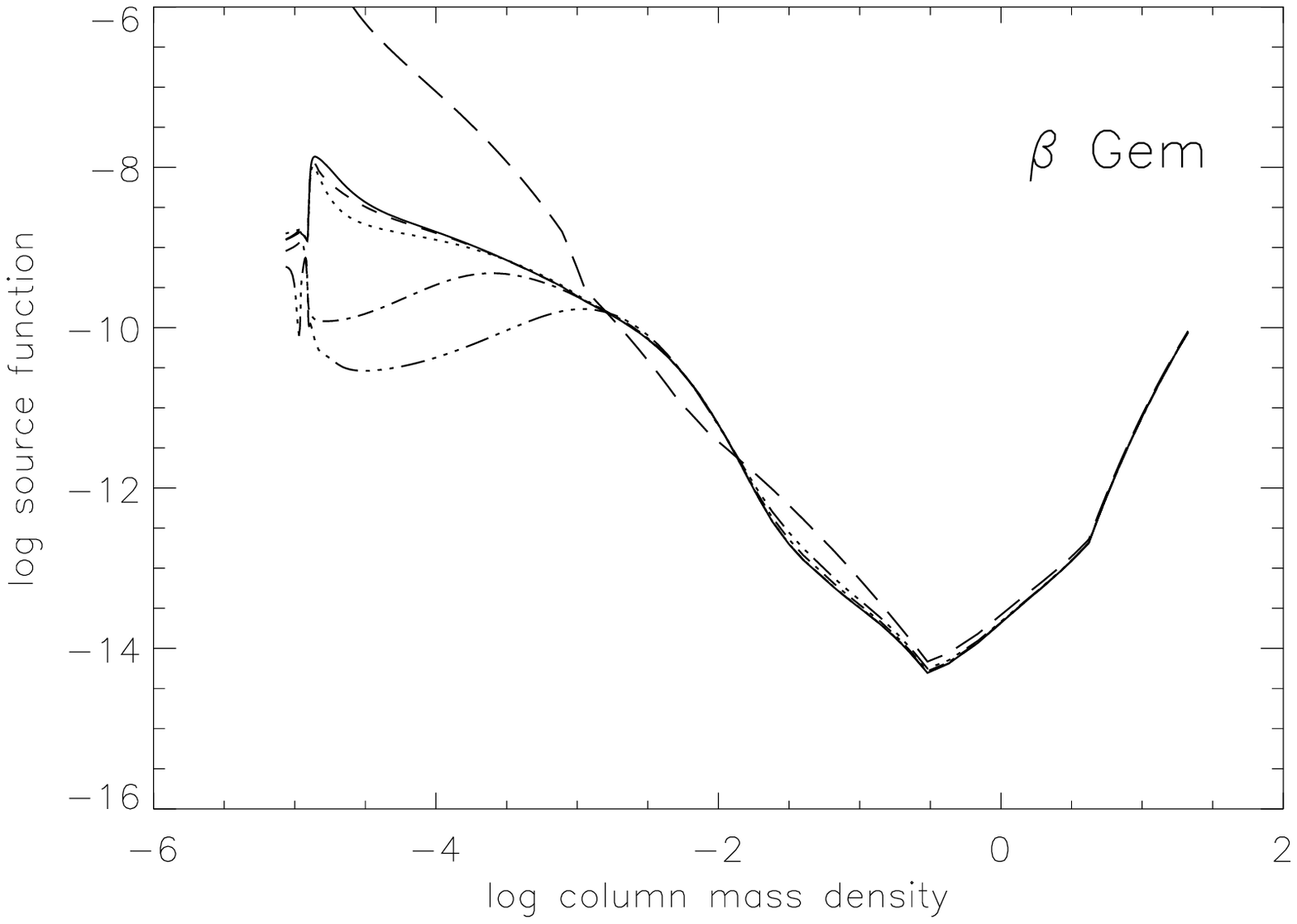,height=7.0cm,width=8.0cm}
\psfig{file=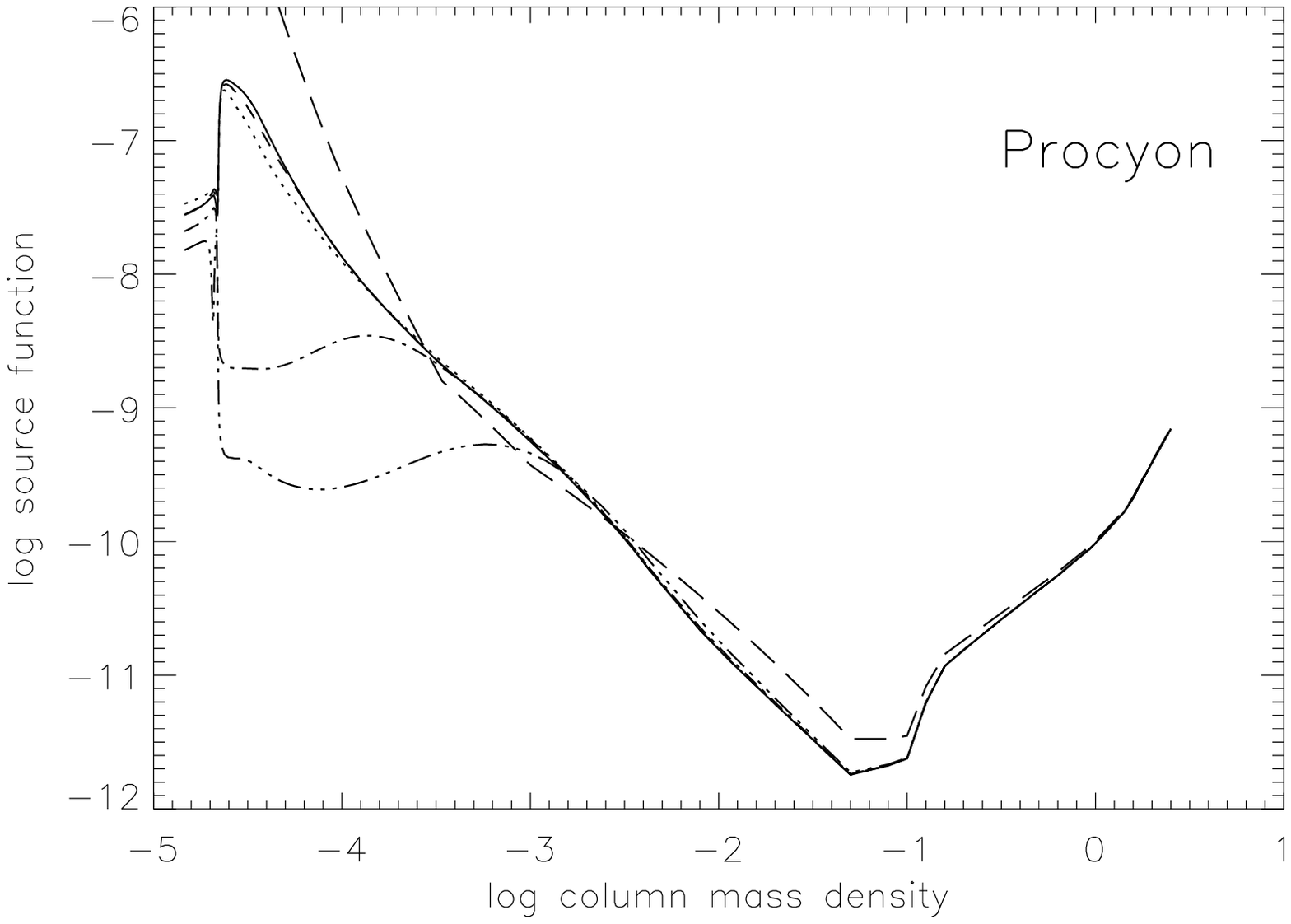,height=7.0cm,width=8.0cm}}} 
\caption{Source functions for Lyman~$\alpha$. In each case the dotted
line is for CRD, the solid line is for PRD ($\Delta \lambda = 0$), the
dashed line is for PRD ($\Delta \lambda = 6 \Delta \lambda_{D}$), the
dash-dot line is for PRD ($\Delta \lambda = 18 \Delta \lambda_{D}$), the
dash-triple-dot line is for PRD ($\Delta \lambda = 35 \Delta \lambda_{D}$) 
and the long-dash line is the Planck function. All source functions
are in ergs~cm$^{-2}$~s$^{-1}$~sr$^{-1}$~Hz$^{-1}$ and column mass density is
in g~cm$^{-2}$.  $\Delta \lambda_{D}$ is the thermal Doppler width at $T_{e}=8
\times 10^{3}$K.} 
\label{atherm}
\end{center}
\end{figure}

Differences between CRD and PRD integrated line fluxes have been shown
by \mylongcite{JudgeTherm} to be related to the line thermalization
depth. In that paper the author shows that in a solar model the Mg~{\sc ii}
$k$-line (which forms in the chromosphere) thermalizes significantly
higher in the chromosphere when using a PRD rather than CRD approach,
while in a model of $\alpha$~Tau, the same line thermalizes, for both
PRD and CRD calculations, not in the chromosphere but in the
photosphere.  Because the effective optical thickness of the line
depends upon the thermalization depth, this means that the discrepancy
between CRD and PRD calculations of the Mg~{\sc ii}~$k$ integrated line flux
is significantly greater in the effectively thick solar model than the
effectively thin $\alpha$~Tau model -- the more effectively thin the
atmosphere the less important are the details of radiative transfer in
calculating the emergent flux.  Applying the same arguments to
Lyman~$\alpha$ would suggest that one should expect the discrepancy
between computed CRD and PRD line fluxes to be significant only when the
atmosphere is effectively thick in Lyman $\alpha$ (i.e. when the
Lyman~$\alpha$ line thermalizes high in the atmosphere).
Figure~\ref{atherm} shows the computed source functions ($S$)
and the Planck function ($B$) for each of the atmospheric models.  The
CRD source function is shown along with the monochromatic source
function for case (d) PRD at four different wavelength detunings
(${\Delta \lambda}/{\Delta \lambda_{D}} = 0, 6, 18, 35$ where $\Delta
\lambda_{D}$ is the thermal Doppler width at $8 \times 10^{3}$K).
It can be seen that thermalization of Lyman $\alpha$ (in the sense 
that $S$ does not significantly deviate from  $B$) occurs in the 
chromosphere in each of the models and there is not a strong variation 
of thermalization depth with surface gravity. This difference between 
the thermalization properties of Lyman $\alpha$ and the Mg~{\sc ii} line 
is due to the different photon destruction processes which thermalize the 
lines. In Mg~{\sc ii} the line is thermalized by electron collisional 
de-excitation and since this process occurs at a rate proportional to 
the electron density the photon destruction rate drops rapidly in the 
outer atmosphere. This means that the source function can deviate 
significantly from the Planck function above a certain level in the 
atmosphere where the electron density has dropped low enough that 
collisional de-excitation can no longer 
thermalize the line. Since the value of the electron density is lower 
in lower gravity stars this decoupling of the source and Planck functions 
will occur sooner (i.e. deeper in the atmosphere) in lower gravity stars. 
In Lyman~$\alpha$, collisional de-excitation is a less important 
thermalization process that photoionization of $n=2$ by the Balmer 
continuum. Unlike the collisional de-excitation rate, the 
photoionization rate does not drop rapidly in the outer atmosphere 
and so this process is effective in preventing a substantial 
deviation of $S$ from $B$ until the high chromosphere where the 
thermodynamic properties of the atmosphere vary sufficiently rapidly 
that the photoionization process is no longer able to couple $S$ to 
the local value of $B$. Thus, Lyman $\alpha$ is not effectively thin 
in any of the stars considered here and so radiative transfer effects 
are important in the calculation of both the line profile and the line 
flux in all the models.

\begin{figure}
\begin{center}
\centerline{\vbox{
\psfig{file=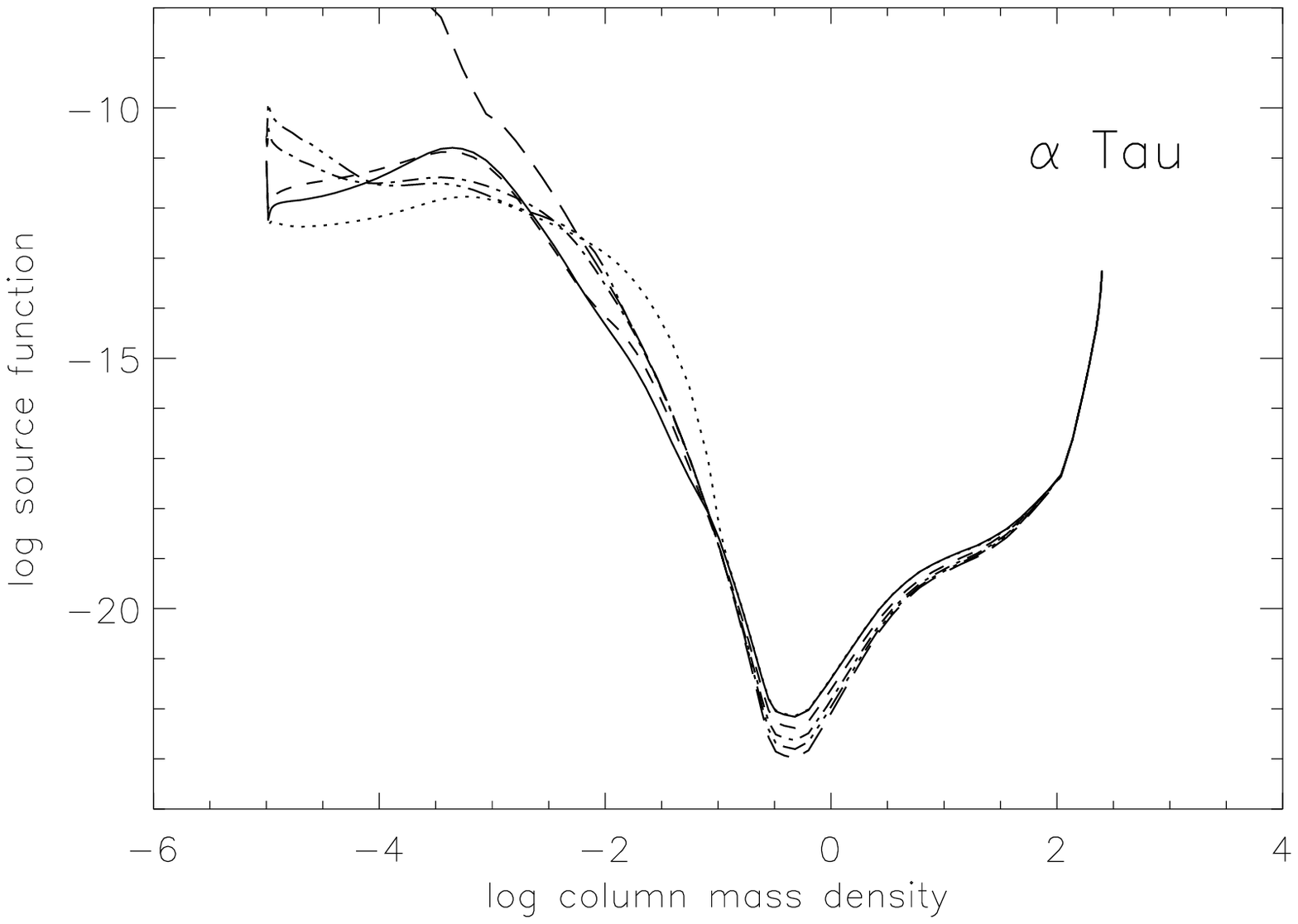,height=7.0cm,width=8.0cm}
\psfig{file=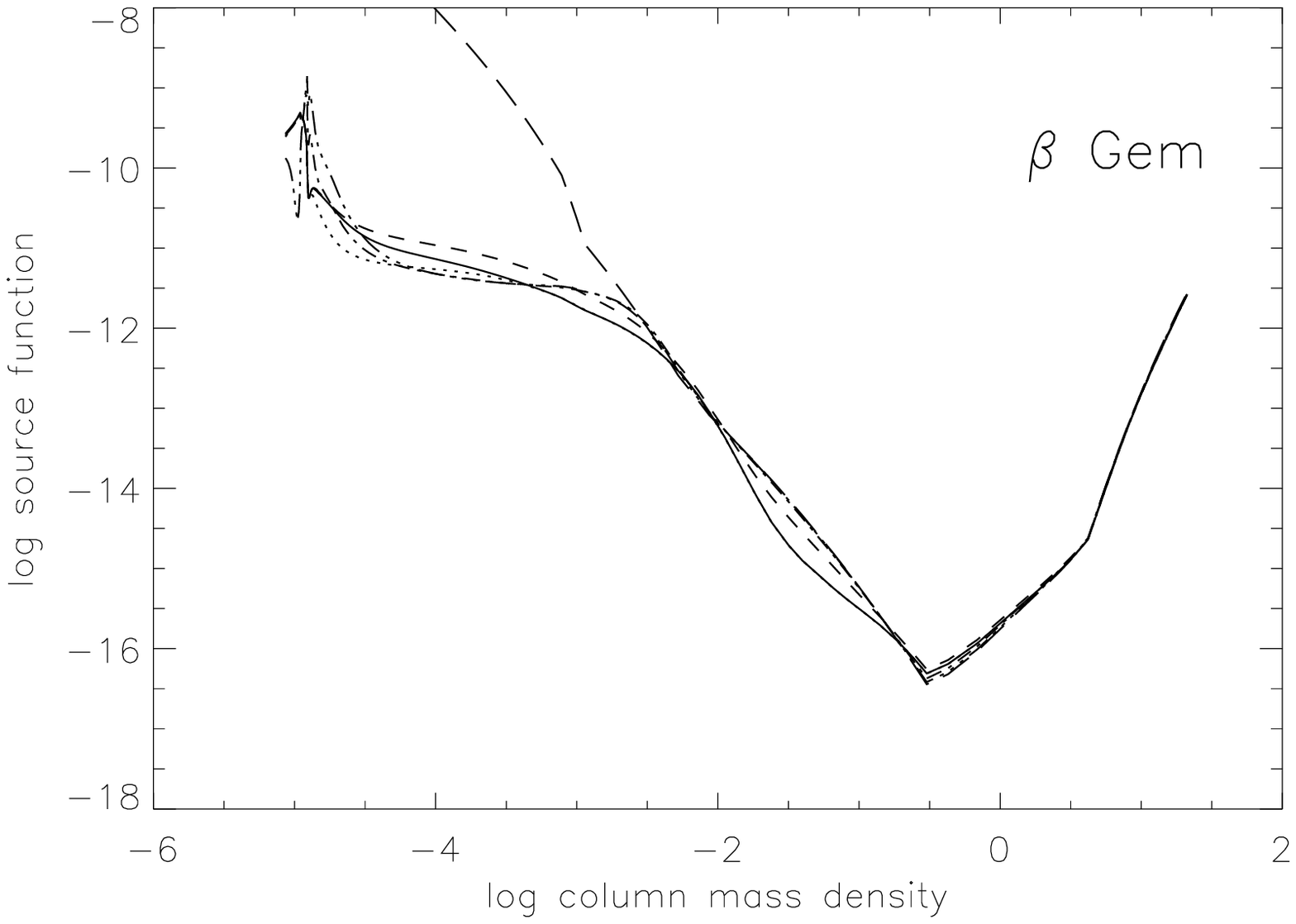,height=7.0cm,width=8.0cm}
\psfig{file=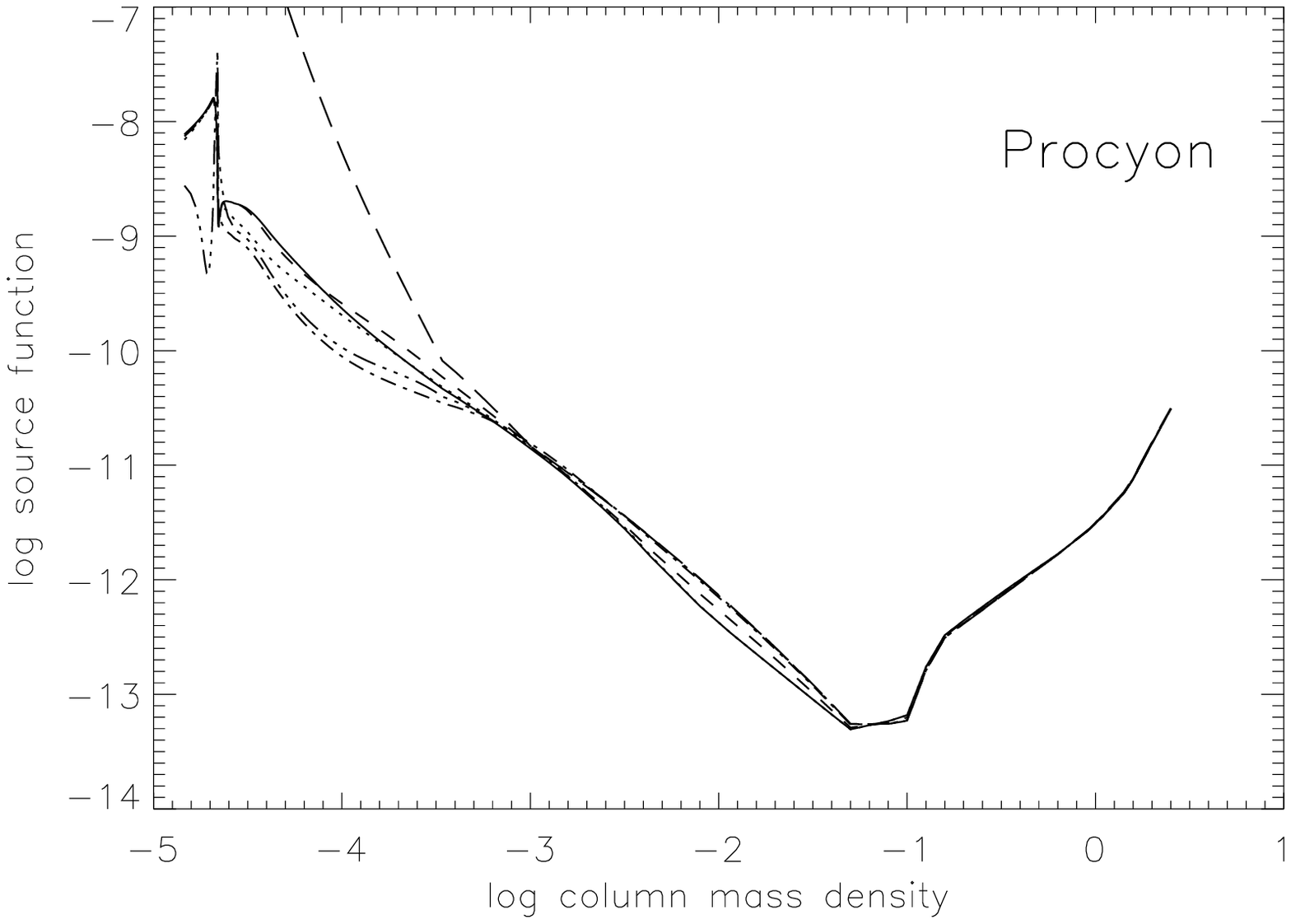,height=7.0cm,width=8.0cm}}} 
\caption{Source functions for Lyman~$\beta$.  In each case the dotted
line is for CRD, the solid line is for PRD ($\Delta \lambda = 0$), the
dashed line is for PRD ($\Delta \lambda = 6 \Delta \lambda_{D}$), the
dash-dot line is for PRD ($\Delta \lambda = 18 \Delta \lambda_{D}$), the
dash-triple-dot line is for PRD ($\Delta \lambda = 35 \Delta \lambda_{D}$)
 and the long-dash line is the Planck function. All source functions
are in ergs~cm$^{-2}$~s$^{-1}$~sr$^{-1}$~Hz$^{-1}$ and column mass density is 
in g~cm$^{-2}$.  $\Delta \lambda_{D}$ is the thermal Doppler width at $T_{e}=8
\times 10^{3}$K.} 
\label{btherm}
\end{center}
\end{figure}

The integrated line flux for Lyman~$\beta$ using CRD agrees with the
PRD flux to within 5 per cent for Procyon and 25 per cent for
$\beta$~Gem, significantly closer agreement than was found for
Lyman~$\alpha$. In $\alpha$~Tau the difference between CRD and PRD
fluxes is somewhat larger (factor of 3.3), but still less 
than the difference that was found in Lyman~$\alpha$.  Figure~\ref{btherm} 
shows the
source functions and Planck function for Lyman~$\beta$. It can be seen 
that, like Lyman $\alpha$, Lyman $\beta$ thermalizes in the high 
chromosphere in each of the models and so this line is not effectively 
thin in any of the models and radiative transfer effects will be 
important, not only in the calculation of the line profile but also in 
the calculation of the total flux. The fact that, despite the Lyman $\beta$ 
line being effectively thick, PRD and CRD calculations predict line fluxes 
that agree substantially better than those found for Lyman $\alpha$ is a 
strong indication of how much less important PRD effects are in 
Lyman $\beta$ than Lyman $\alpha$, however it must be stressed that the 
effects of PRD on the Lyman $\beta$ flux are still significant, 
particularly in the low gravity stars.

\subsection{Implications for Modelling}
\label{nhp}

\begin{figure}
\begin{center}
\centerline{\vbox{
\psfig{file=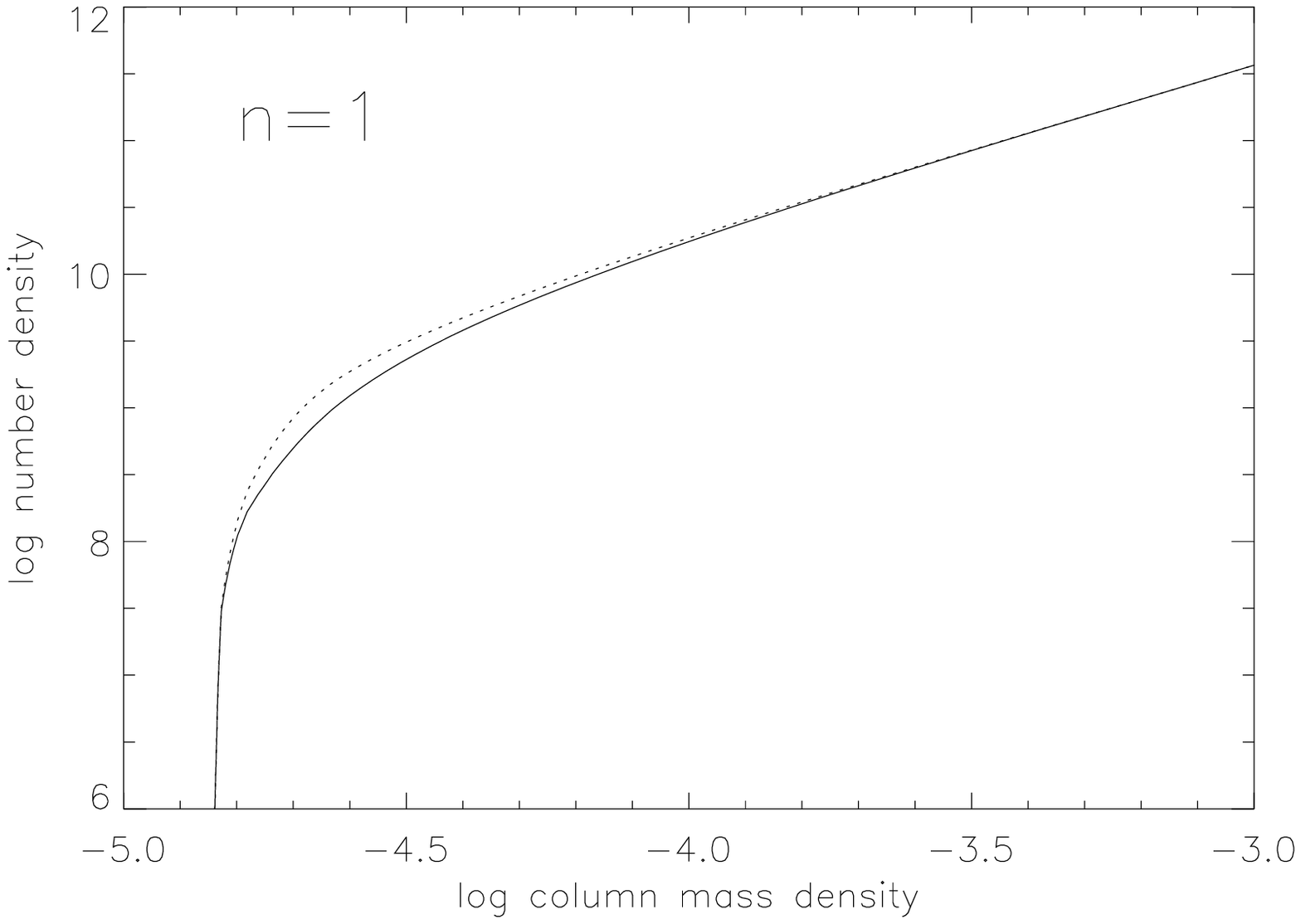,height=5.4cm,width=8cm}
\psfig{file=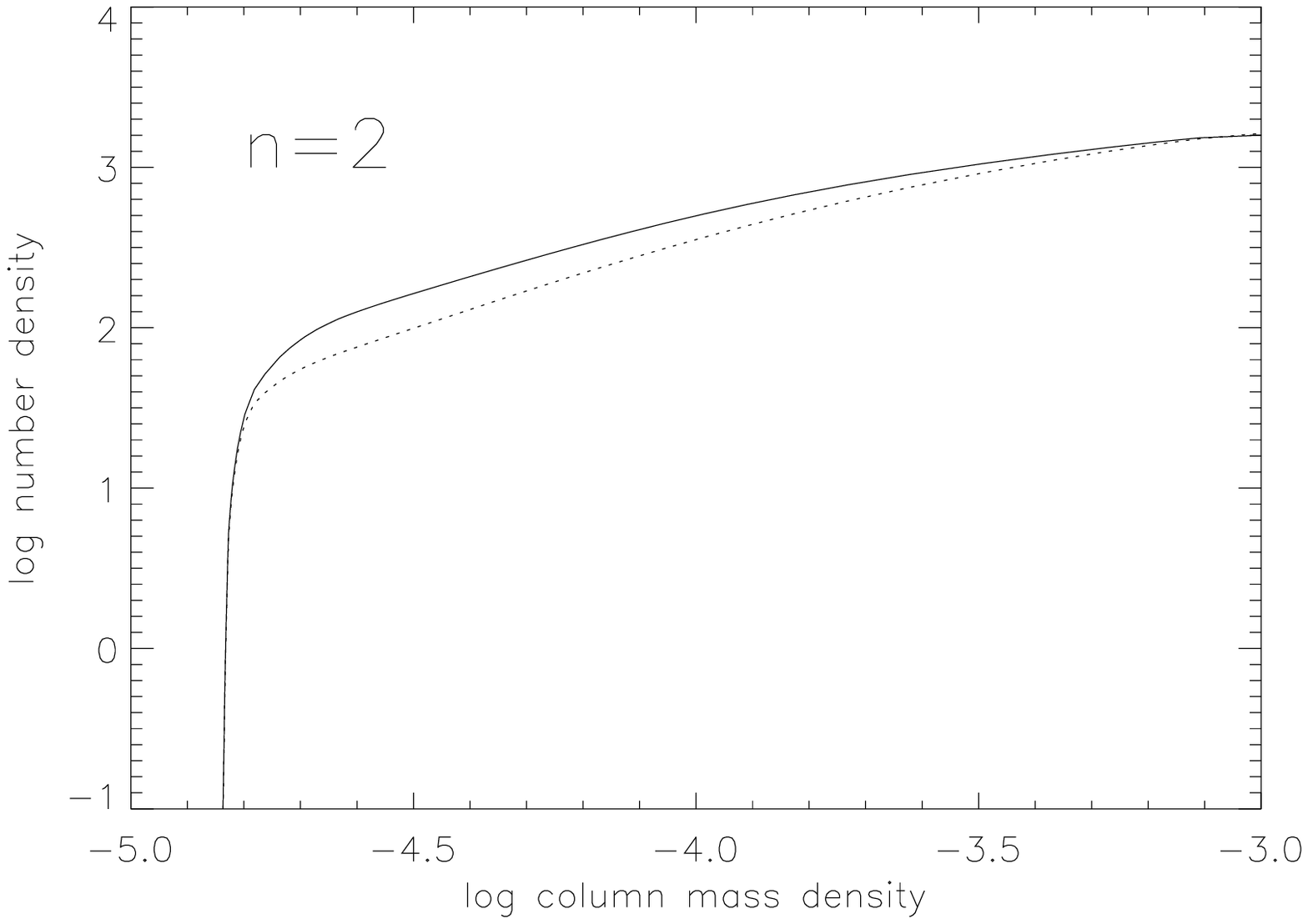,height=5.4cm,width=8cm}
\psfig{file=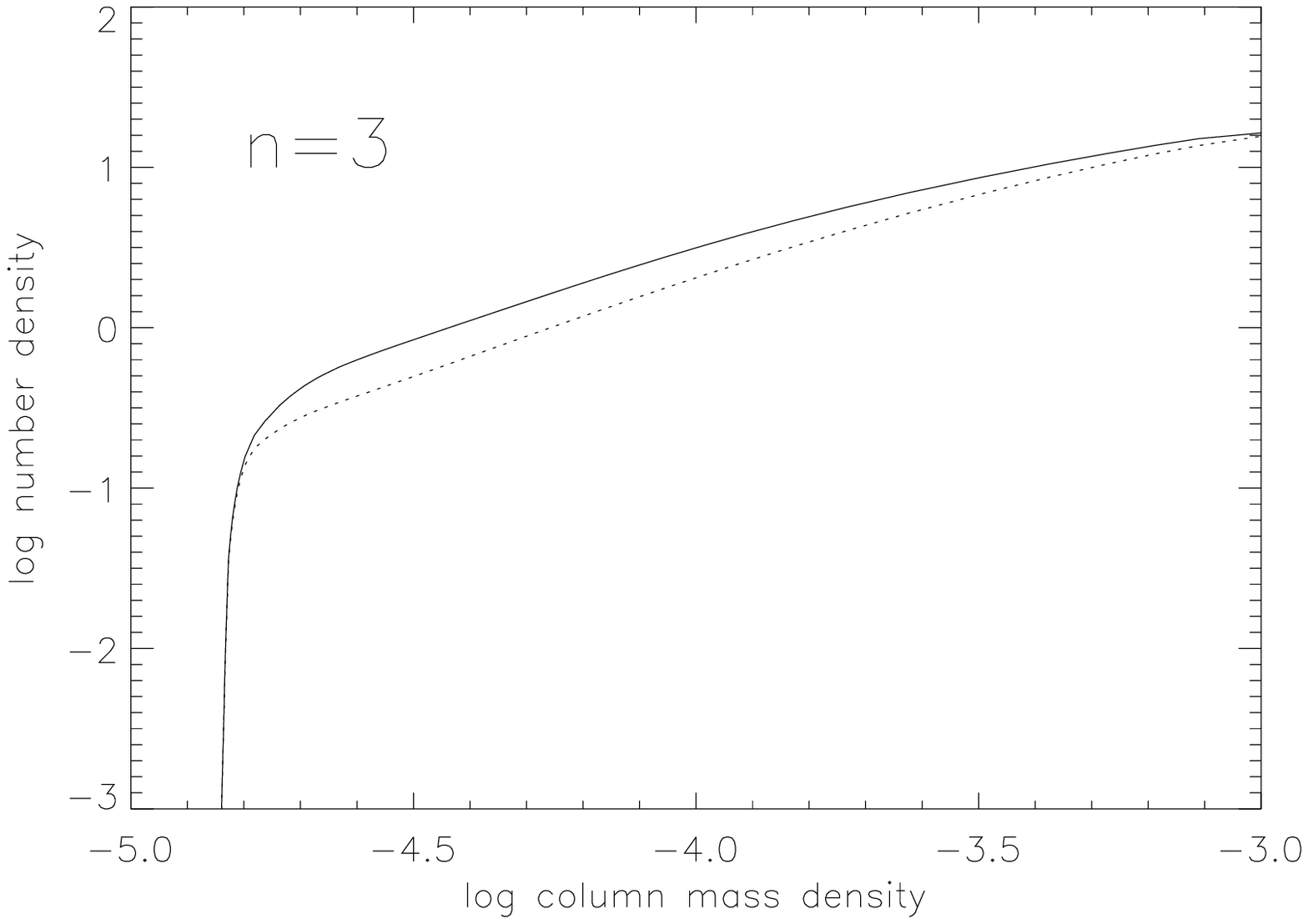,height=5.4cm,width=8cm}
\psfig{file=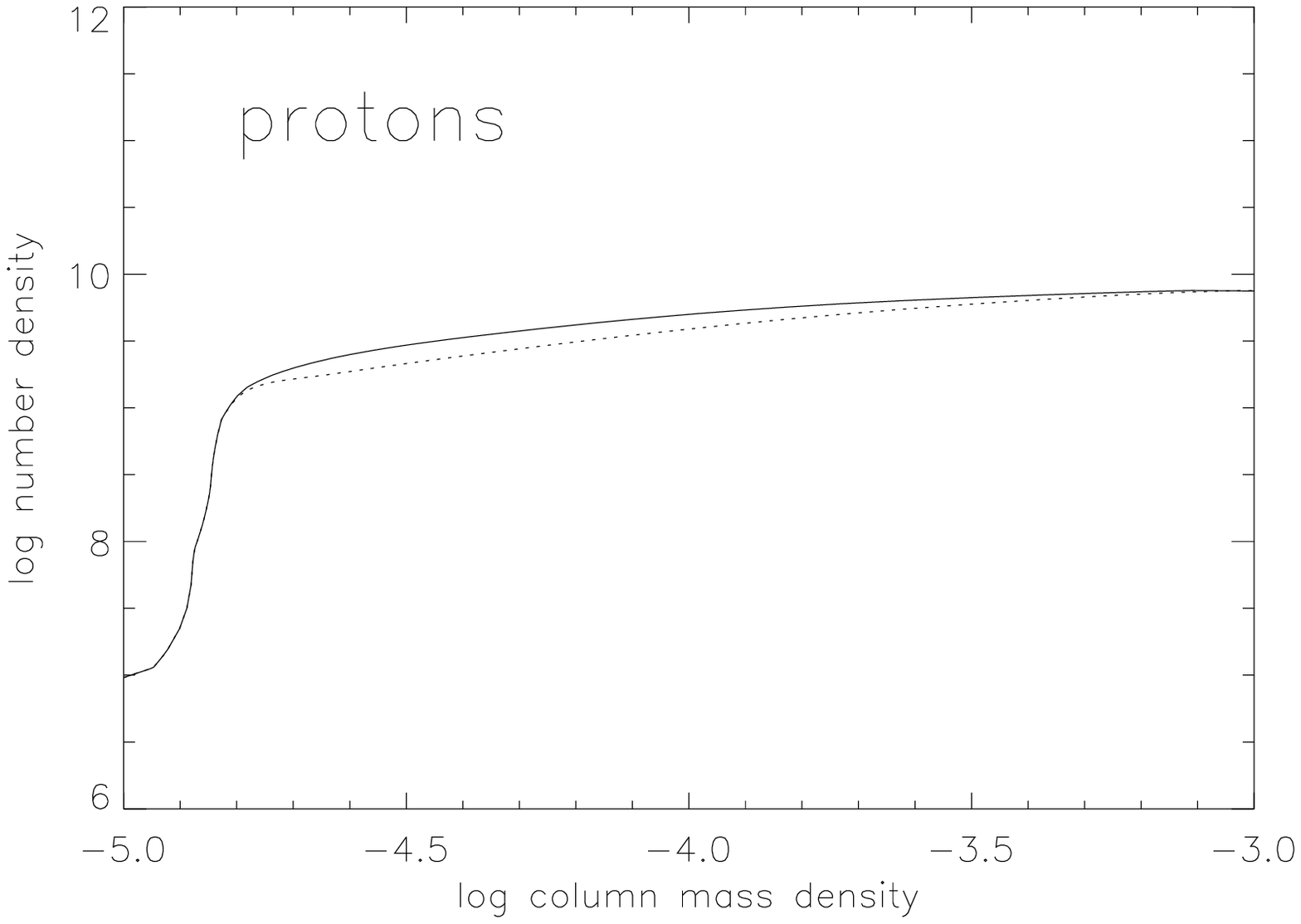,height=5.4cm,width=8cm}}}
\caption{Calculated hydrogen level populations for $\beta$~Gem. The
solid line is case (d) PRD and the dotted line is CRD. Population
number densities are in cm$^{-3}$ and column mass density is in
g~cm$^{-2}$.}
\label{levelpop}
\end{center}
\end{figure}

The calculations discussed above were all carried out using fixed
atmospheric models. However, this process is not strictly
self-consistent. It is known that using a PRD (rather than CRD)
treatment alters the level populations of the hydrogen atom
(\mycite{HL}, \mycite{LJ92}). Figure~\ref{levelpop} shows the level
populations calculated for the four different redistribution cases
using the $\beta$~Gem atmospheric model. The variation in the level
populations is similar to that found by \mylongcite{HL}: the $n=2$ and
$n=3$ level populations rise because of the increased trapping of
photons in the Lyman~$\alpha$ line (as described in Section~4.1). An 
increased $n=2$ population means that
the proton population also rises because the primary ionization
mechanism is photoionization from $n=2$. These increases in the $n=2$, $n=3$ 
and proton populations are balances by a decrease in the $n=1$ ground state 
population. Physically, a significant
increase in the proton density carries with it an increase in the
electron density and so to hold the electron density constant (at a
value calculated from the hydrostatic equilibrium assumption in CRD)
is not acceptable. Therefore, it is necessary to use a PRD treatment
of hydrogen when constructing a model atmosphere in order that PRD
calculation results can be consistent with the assumption of
hydrostatic equilibrium.

\begin{figure}
\begin{center}
\centerline{\vbox{
\psfig{file=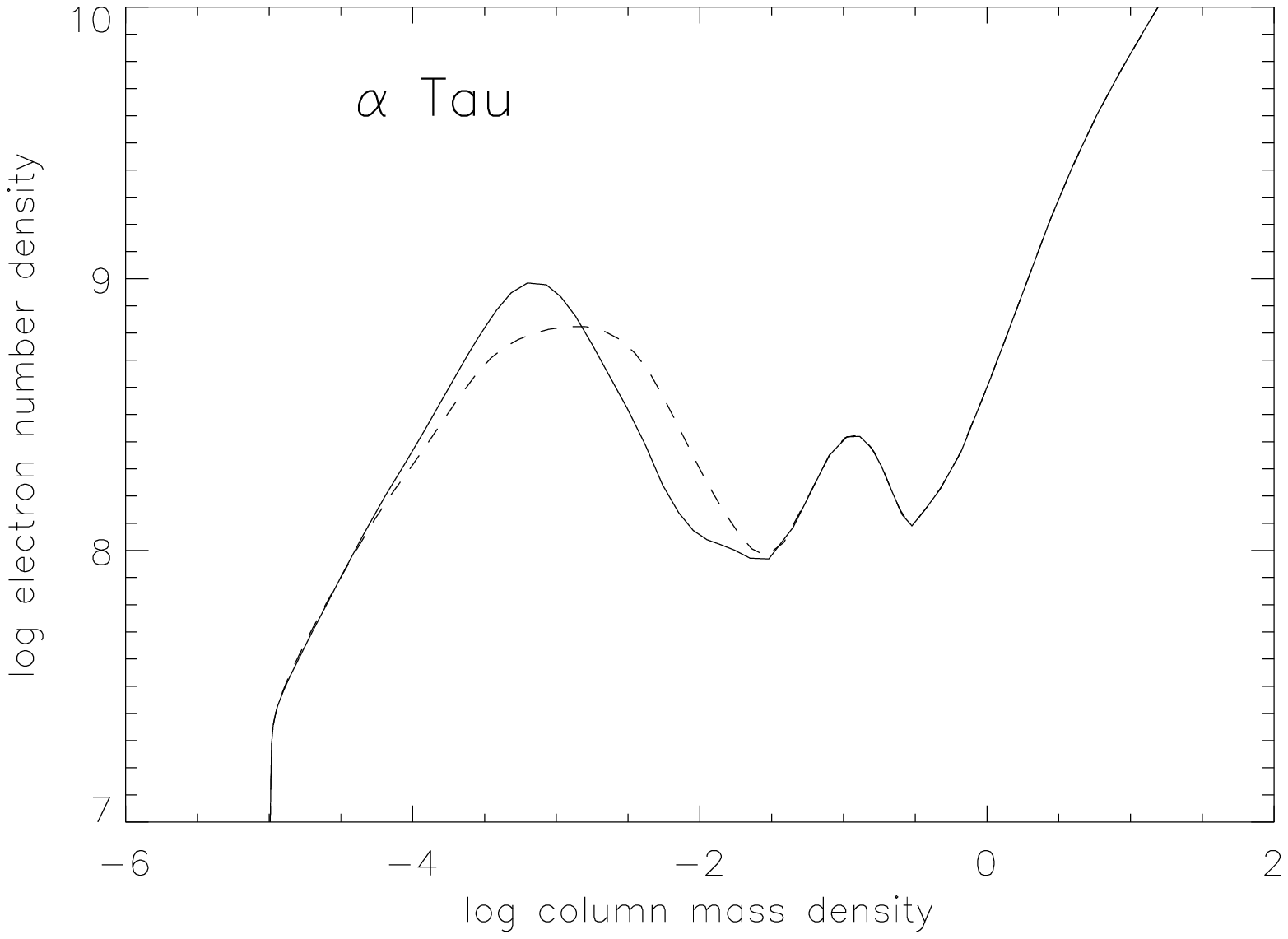,height=7.0cm,width=8.0cm}
\psfig{file=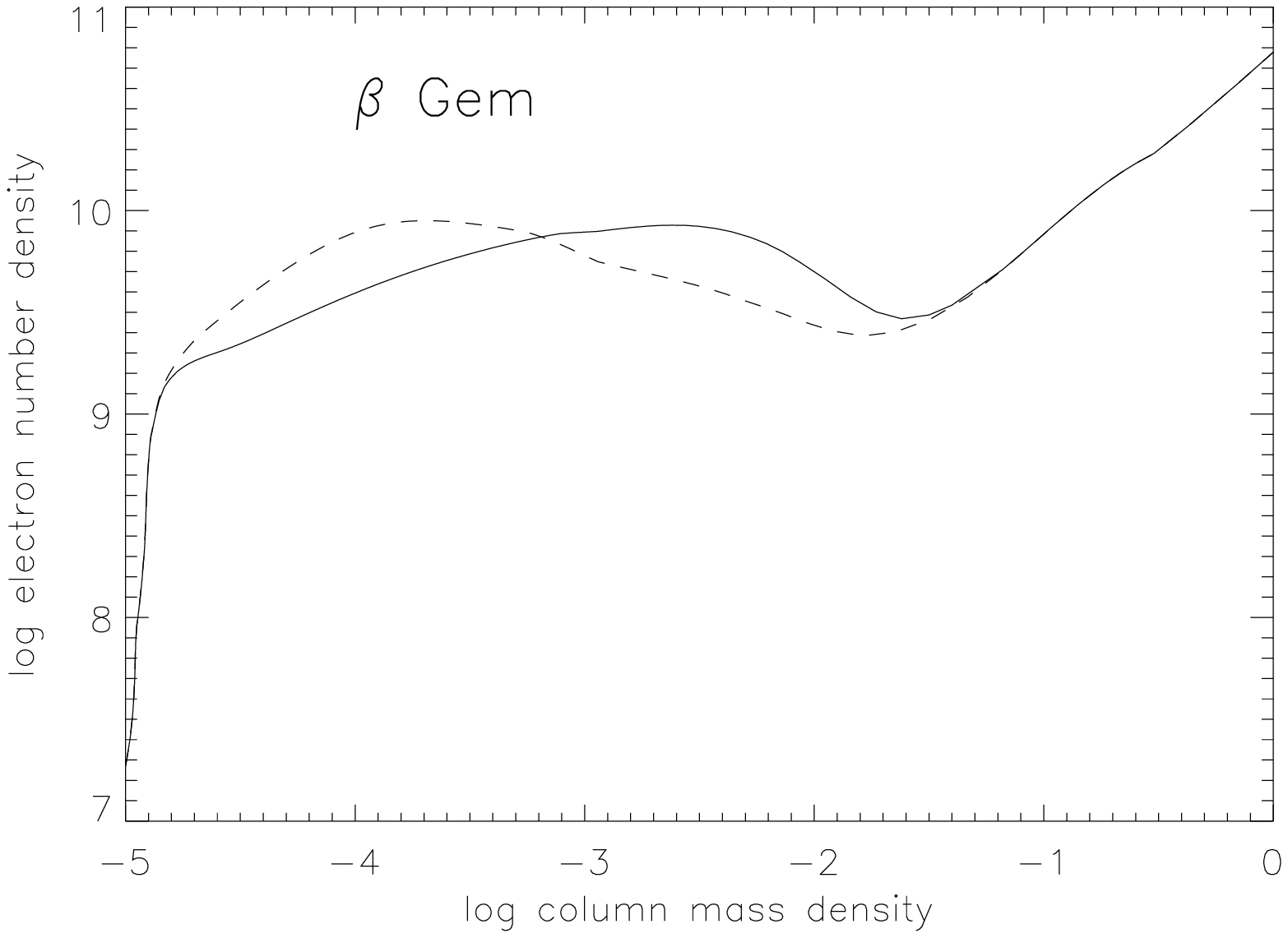,height=7.0cm,width=8.0cm}
\psfig{file=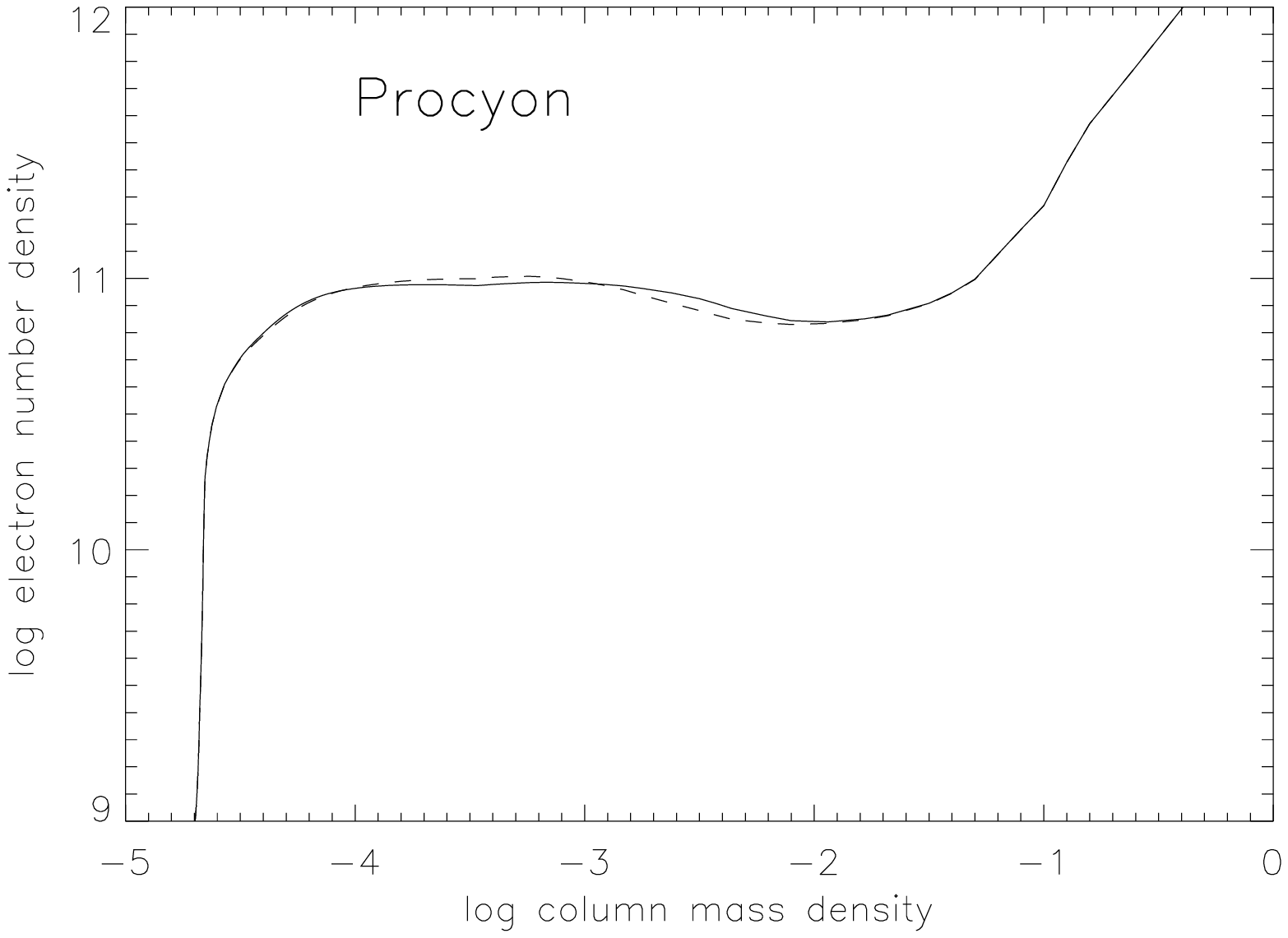,height=7.0cm,width=8.0cm}}}
\caption{Electron densities from hydrostatic equilibrium
calculations. The solid lines are from the original (CRD) models and
the broken lines are from models constructed based on hydrostatic
equilibrium models computed using a PRD treatment of hydrogen. The
electron number density is in cm$^{-3}$ and the column mass density is in
g~cm$^{-2}$.}
\label{ne}
\end{center}
\end{figure}

For this reason, each of the atmospheric models has been recomputed
using a PRD treatment of hydrogen in the hydrostatic equilibrium
integrations. For $\alpha$~Tau, the original model has simply been
recomputed in hydrostatic equilibrium with PRD. This means that the
new model of $\alpha$~Tau is no longer strictly consistent with the
emission measure distribution from which it was derived by
\mylongcite{Mcmurry99}. The models of $\beta$~Gem and Procyon have been
recomputed in hydrostatic equilibrium with PRD and have also been
iterated to consistency with the emission measure distributions from
which they were derived. The only difference between the three new
PRD-based models and the original three models is in the electron
density. Figure~\ref{ne} shows the change in the electron density in
each model. $\beta$~Gem shows the largest differences, up to a factor
of three between the electron density in CRD and PRD calculations. As
would be expected, these differences occur in the chromosphere and low
transition region, where the Lyman lines are forming. Procyon shows
the smallest differences in the electron density of not much more that
20 per cent. $\alpha$~Tau shows significant differences between the CRD and
PRD calculations, in a similar way to $\beta$~Gem, but the maximum
change in $\alpha$~Tau is only about a factor of two. It may seem
surprising that the lower gravity star $\alpha$~Tau shows smaller
changes than $\beta$~Gem, but it must be borne in mind that the direct
comparison of these two models is not valid: the original $\alpha$~Tau
model is not a true CRD model but was constructed by
\mylongcite{Mcmurry99} using an approximate PRD method. 

Figure~\ref{ne} makes clear that a PRD treatment of hydrogen is
important in constructing atmospheric models, particularly but not
exclusively, for low gravity stars. In, for example, the model of
$\beta$~Gem, the electron density in the chromosphere is very
significantly different from that predicted by the CRD treatment and
it will lead to substantially different fluxes of collisionally
excited lines of all elements.  It is not within the scope of this
paper to give a detailed discussion of the atmospheric emission lines
of elements other than hydrogen, but the importance of the PRD
treatment of hydrogen on these lines will be discussed in subsequent
papers. In particular the collisionally formed lines of C~{\sc ii} and the
fluorescent lines of O~{\sc i} (pumped by Lyman~$\beta$) will be of
interest. Table~\ref{bronzefluxes} gives the integrated fluxes of the
Lyman~$\alpha$ and Lyman~$\beta$ lines calculated using the full case
(d) PRD with the new PRD-based models. If these fluxes are compared
with the case (d) results in Table~\ref{fluxcal} it can be seen that
the difference in the fluxes reflects the difference in the electron
density: $\beta$~Gem shows a very significant difference, $\alpha$~Tau
shows a significant difference but Procyon shows very little
difference.

\begin{table}
\begin{center}

\caption{Calculated integrated line fluxes from the four new PRD-based 
models with
case (d) PRD treatment of hydrogen. All fluxes are at stellar surface
in ergs~cm$^{-2}$~s$^{-1}$.}
\begin{tabular}{|c|c|c|c|c|c|} \hline
Transition &$\alpha$~Tau & $\beta$~Gem & Procyon \\ \hline
Lyman~$\alpha$ & $1.23 \times 10^{4}$ & $3.20 \times 10^{4}$ & $4.03
\times 10^{5}$ \\ 

Lyman~$\beta$  & $3.79 \times 10^{1}$ & $3.44 \times 10^{2}$ & $7.50
\times 10^{3}$ \\ \hline
\label{bronzefluxes}

\end{tabular}
\end{center}
\end{table}

\subsection{Comparison with Observations}

\begin{figure}
\begin{center}
\centerline{\vbox{
\psfig{file=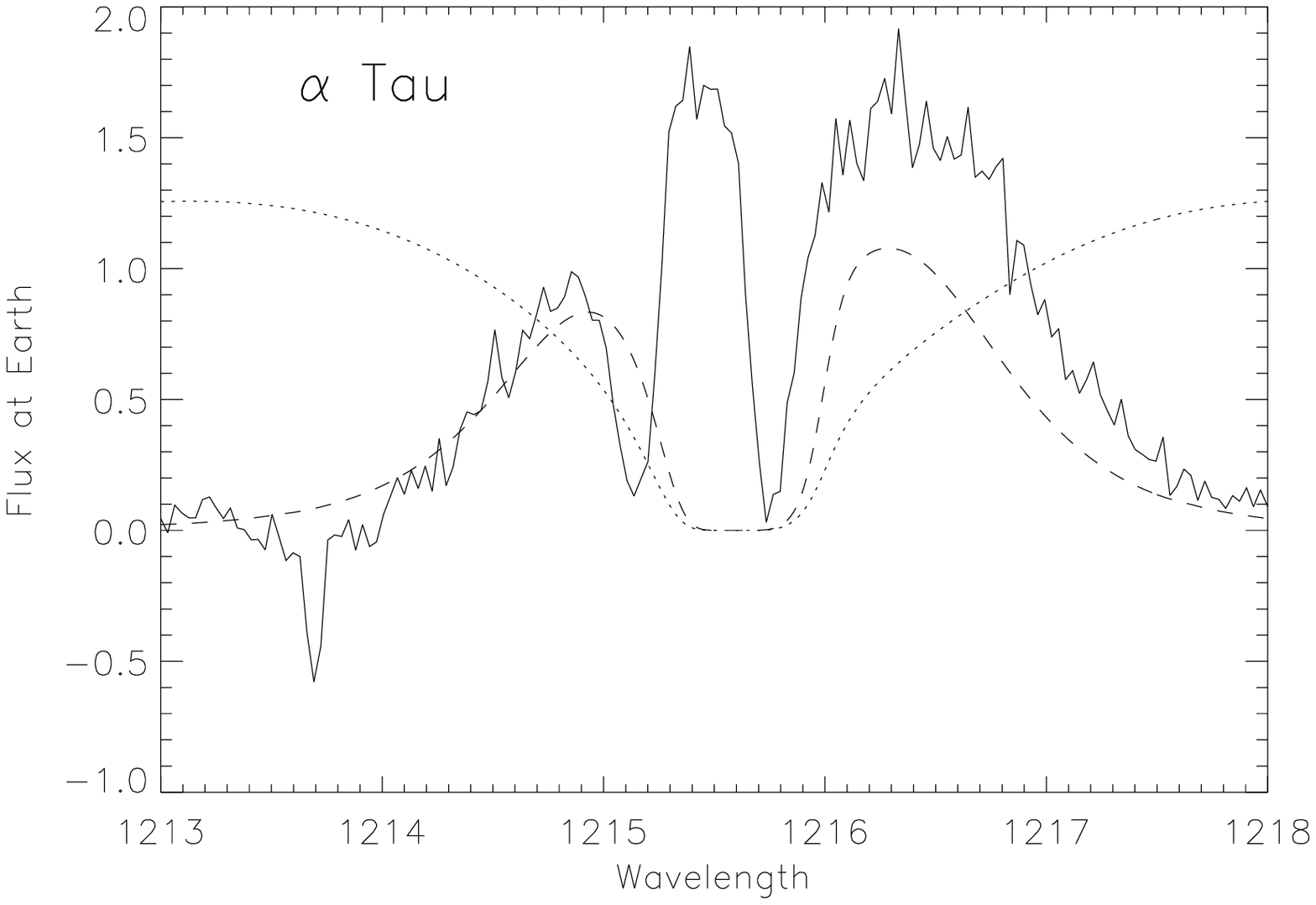,height=7.0cm,width=8.0cm}
\psfig{file=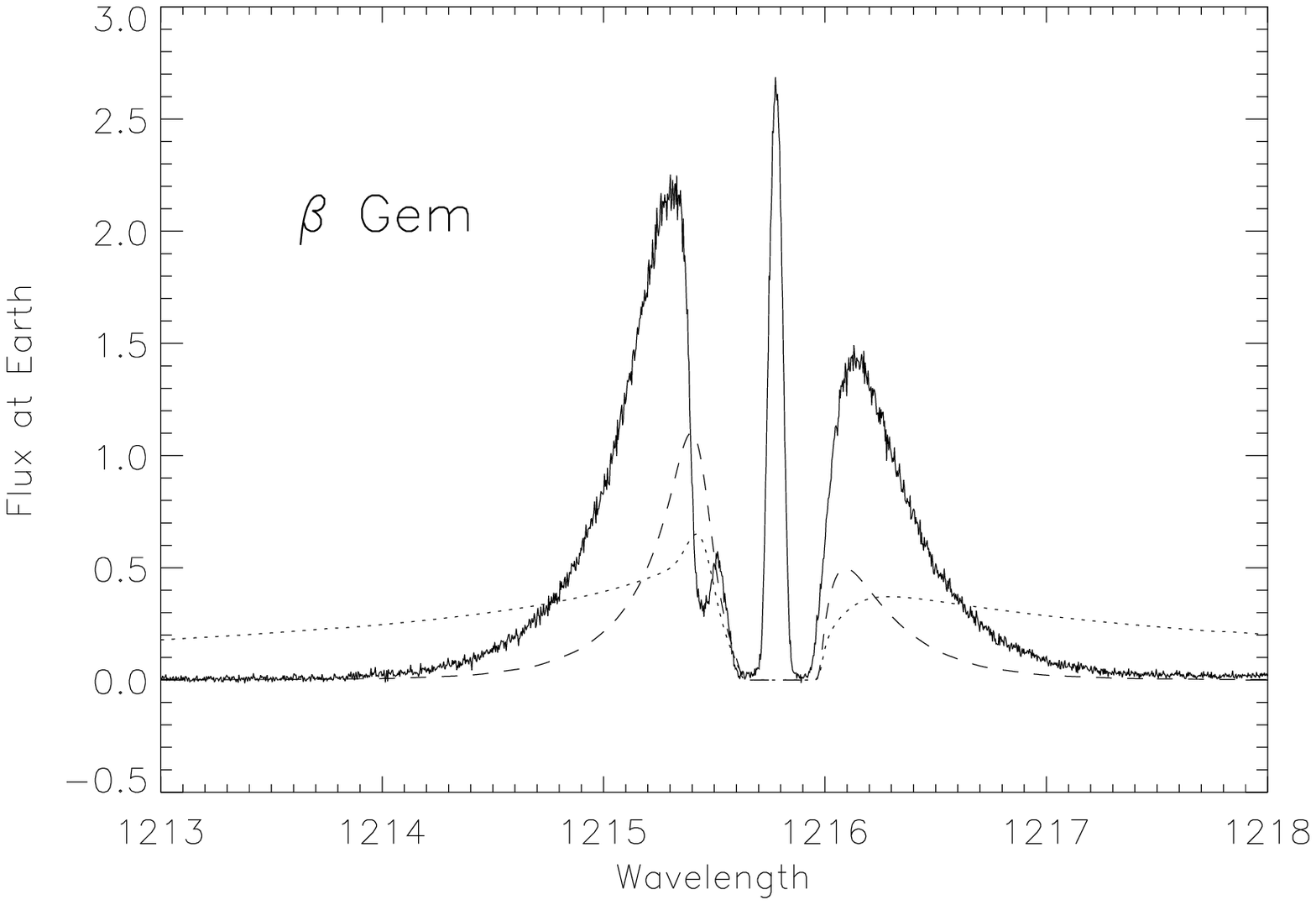,height=7.0cm,width=8.0cm}
\psfig{file=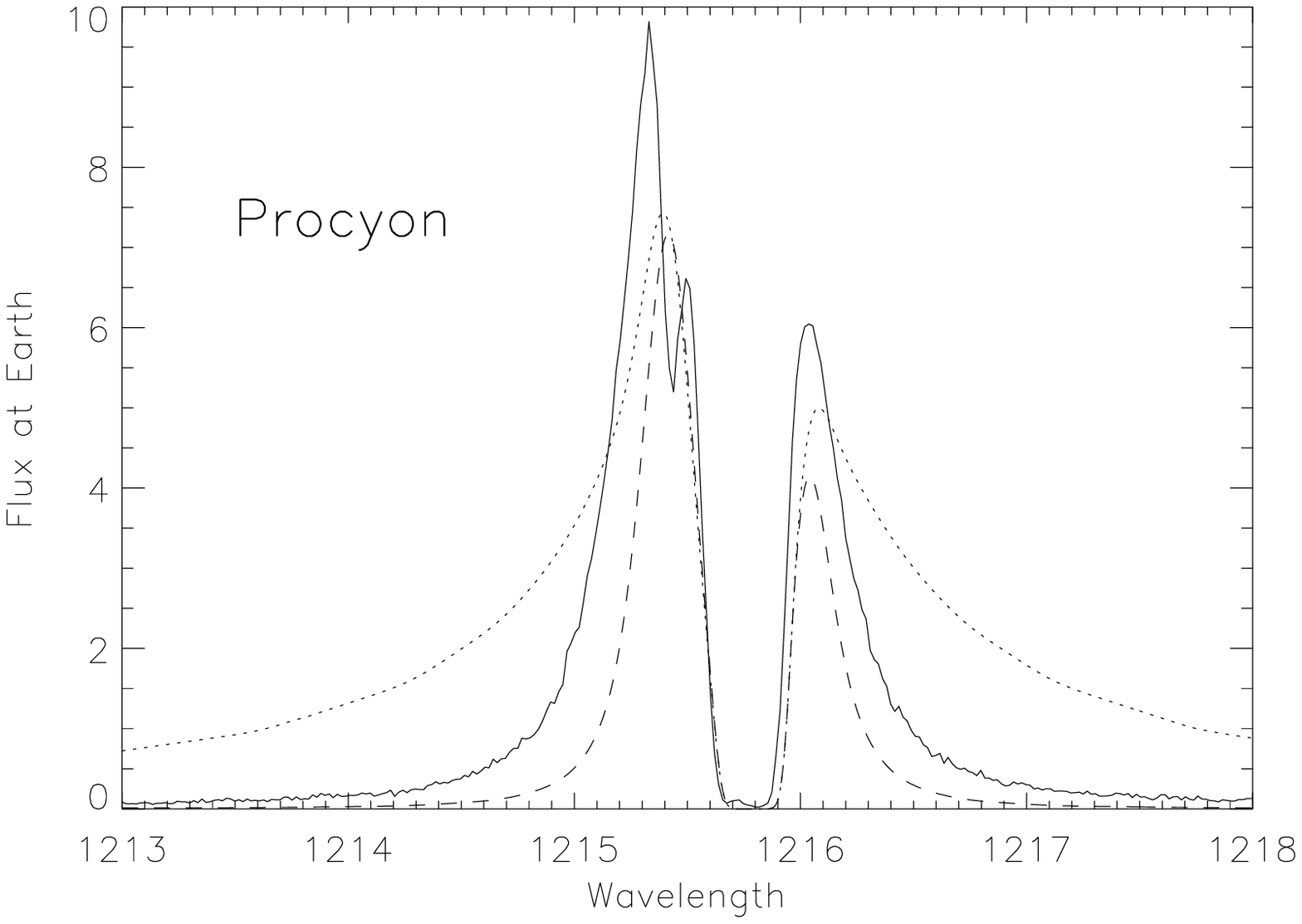,height=7.0cm,width=8.0cm}}}
\caption{Observed and calculated Lyman~$\alpha$ profiles. The solid
line is the observed flux, the dotted line is the calculated flux using CRD 
and the dashed line is the calculated flux using PRD, in units
of $10^{-11}$~ergs~cm$^{-2}$~s$^{-1}$~\AA$^{-1}$. The wavelength is in
\AA. The central emission feature in $\alpha$~Tau and $\beta$~Gem is due to 
geocoronal emission.}
\label{obs}
\end{center}
\end{figure}

The work discussed in the previous sections has led to the conclusion
that,  for a given atmospheric model, predictions based on a CRD
treatment of hydrogen differ from those based on a PRD treatment in
three ways. First and foremost, CRD leads to predicted line profiles
for Lyman~$\alpha$ that have substantially stronger wings (and hence
greater integrated line flux) than the more satisfactory PRD
treatment. PRD also predicts significantly different profiles for 
Lyman~$\beta$ in low gravity stars -- this effect should be observable 
in future observations of late-type stars with the {\it FUSE} instrument. 
Secondly, because the line profiles are different, there may
be differences in the properties of lines that are excited by
fluorescence in the Lyman lines, depending on the wavelength at which
lines are pumped.  Thirdly, because a PRD treatment leads to a
different chromospheric electron density there will be a change in the
strength of collisionally excited chromospheric/transition region
lines (such as the C~{\sc ii} lines at around 1335~\AA). The secondary
effects on the lines of other atomic species will not be discussed
here but will be addressed in subsequent papers on the details of the
atmospheric models. The most easily observed consequence of PRD (as
opposed to CRD) is the significant relative reduction of the line wing
intensity in Lyman~$\alpha$ compared to the line
core. Figure~\ref{obs} show observed Lyman~$\alpha$ profiles and the
synthetic profiles calculated using the original models with case (a) CRD, 
and also the new PRD models treated with full case (d) PRD. The observed
profile for $\alpha$~Tau is from the {\it IUE} high dispersion exposure
SWP6679 (C. Jordan (PI)). The observations for Procyon and $\beta$~Gem are 
from GHRS
exposures Z17X0303M (J. Linsky (PI)) and Z2SI0406T (R. Henry (PI)) 
respectively.

\begin{table}
\begin{center}
\caption{Parameters used to correct calculated line profiles for
interstellar H~{\sc i} absorption and macroturbulence. For $\beta$~Gem and
Procyon there are two separate components to the interstellar
absorption.}

\begin{tabular}{|c|c|c|} \hline
Star & Parameter & Value \\ \hline

$\alpha$~Tau & $\xi_{\mbox{\scriptsize macro}}$ (km~s$^{-1}$) & 17  \\
 & $\xi_{\mbox{\scriptsize micro}}$ (km~s$^{-1}$) & 13  \\
& $N_{H}$ (cm$^{-2})$ & $5 \times 10^{18}$$^{a}$ \\ & $b$
 (km~s$^{-1}$) & 11$^{b}$  \\ & $v_{0} ($km~s$^{-1}$) & - 30$^{c}$  \\
 \hline

$\beta$~Gem & $\xi_{\mbox{\scriptsize macro}}$ (km~s$^{-1}$) & 16 \\ 
& $\xi_{\mbox{\scriptsize micro}}$ (km~s$^{-1}$) & 12 \\ 
& $N_{H}$ (cm$^{-2})$ & $1.15\times 10^{18}$/$6.82\times 10^{17}$$^{d}$
 \\ 
& $b$ (km~s$^{-1})$ & 12.33/11.02$^{d}$ \\ & $v_{0}$ (km~s$^{-1})$
 & 21.9/33.0$^{d}$  \\ \hline

Procyon & $\xi_{\mbox{\scriptsize macro}}$ (km~s$^{-1}$) & 29   \\ 
& $\xi_{\mbox{\scriptsize micro}}$ (km~s$^{-1}$) & 14 \\
&$N_{H}$ (cm$^{-2}$) & $7.5\times 10^{17}$/$4.0\times 10^{17}$$^{e}$
 \\ 
& $b$ (km~s$^{-1}$) & 10.78/10.78$^{e}$ \\ & $v_{0}$ (km~s$^{-1}$)
 & 20.5/21.0$^{e}$\\ \hline
\label{MacAbs}

\end{tabular}
\end{center}

{$^{a}$ \mylongcite{Mclintock}

$^{b}$ assumed

$^{c}$ based on Mg~{\sc ii} $k$-line (see text)

$^{d}$ \mylongcite{Dring}

$^{e}$ \mylongcite{Proabs}}

\end{table}

The atmospheric models contain a microturbulent velocity 
($\xi_{\mbox{\scriptsize micro}}$) which is taken into account in the 
radiative transfer calculations. In order to correctly match the observed 
widths of optically thin lines, computed profiles generally have to be 
broadened by convolution with a Gaussian profile whose width corresponds 
to a most probably macroturbulent speed $\xi_{\mbox{\scriptsize macro}}$. 
Macroturbulence is interpreted as large scale turbulence in the stellar
atmosphere.
The values of $\xi_{\mbox{\scriptsize macro}}$ which have been used are 
shown (along with the microturbulent 
velocity $\xi_{\mbox{\scriptsize micro}}$ at $T_{e} = 10^{4}$K) in
Table~\ref{MacAbs}. These values of $\xi_{\mbox{\scriptsize macro}}$
have been deduced from the widths of optically thin emission lines
that form in similar regions of the atmosphere to the Lyman~$\alpha$
line and will be discussed further in later work regarding the
atmospheric models. The computed profiles have been corrected for 
interstellar absorption. This has been performed using the optical 
depth described by
\mylongcite{Munch} in terms of the hydrogen column density $N_{H}$, the
line width parameter $b$ (which in the current work is taken as
$\sqrt{2}$ times the Doppler dispersion parameter) and the wavelength
shift relative to line centre $v_{0}$. The adopted parameters for the
interstellar absorption are shown in Table~\ref{MacAbs}. For Procyon
and $\beta$~Gem, \mylongcite{Proabs} and \mylongcite{Dring} respectively
have deduced parameters for the interstellar absorption using two
component models. Both components of the H~{\sc i} absorption have been
included here. For $\alpha$~Tau \mylongcite{Mclintock} give only the
column density and so the line width has been taken as a typical value
of 11~km~s$^{-1}$. The velocity of the interstellar component has been 
taken to be the same as the velocity of the interstellar absorption 
component reported in the $\alpha$~Tau Mg~{\sc ii} $k$-line 
by \mylongcite{Robinson98}. For comparison
with the observations, the final $\alpha$~Tau profile has been convolved
with a Gaussian of FWHM$=25$~km~s$^{-1}$ to take the resolution of the
{\it IUE} instrument into account.

It is clear from Figure~\ref{obs} that the PRD line profiles are in
much better agreement with the observed line shapes than are CRD
profiles, a clear indication that a PRD treatment is very important to 
successfully modelling the Lyman $\alpha$ line. The calculated line 
profiles do not agree closely with the observed profiles, but this is 
most likely due to inadequacies in the model atmospheres. In particular 
the total Lyman~$\alpha$
flux is being significantly underpredicted in $\beta$~Gem. 
It is beyond the scope of this paper to recompute the atmospheric models, 
but this will be done in future work. The asymmetry of the observed 
Procyon and $\beta$~Gem
profiles is well accounted for by the effects of interstellar
absorption. The significant asymmetry of the $\alpha$~Tau profile is 
likely to be due to flows in the atmosphere
(or stellar wind) which are not included in the hydrostatic model used
here. The observed profiles also show additional features due to
deuterium absorption (which is seen in the blue wing of the Procyon
and  $\beta$~Gem profiles at $\Delta \lambda = -0.3$\AA ~from line center) 
and geocoronal emission (which is seen as
a tall spike near the centre of the $\beta$~Gem and $\alpha$~Tau
profiles). 

\section{Conclusions}

It has been demonstrated that there are substantial differences
between the results of PRD and CRD calculations of the Lyman lines,
and that these differences are significant in the modelling of stellar
atmospheres. The importance of PRD effects grows in lower surface
gravity stars, but the effects are not negligible in any of the
evolved stars being studied here. The primary effect has been shown to
be a significant reduction in the Lyman~$\alpha$ wing intensity which
cause a major reduction in the computed Lyman~$\alpha$ flux and has
significant repercussions on the atmospheric structure because of the
way in which it modifies the ionization balance in hydrogen. The
differences in Lyman~$\beta$ are less significant and it appears that
for lower gravity stars (such as Procyon) this line may be modelled
reasonably well with a CRD assumption. However, it has also been
demonstrated that this is not always the case: the differences in
$\alpha$~Tau are substantial.

The significance of these results goes beyond simply predicting the
Lyman line profiles and fluxes. Because Lyman~$\alpha$ is such an
important radiative energy loss mechanism, its behaviour has
repercussions for the understanding of chromospheric heating. The
treatment of this line has the power to change the structure deduced
for the atmosphere, causing a significant change in the electron
density which will manifest itself in the predicted fluxes of
collisionally excited emission lines. The issue of fluorescence is
also significant and, depending on the wavelength of the pumped lines,
the differences between the PRD and CRD profiles could lead to
different predicted fluxes of the fluorescent lines. This is important
since the lowest gravity stars show the richest spectra of fluorescent
lines and it is these stars which show the strongest PRD effects.

The atmospheric models that have been employed as test cases in this
paper are preliminary (it is clear from the comparison in the previous
section that they do not, at present, entirely reproduce the
observations) but these models provide a useful set of realistic test
cases to examine the importance of PRD in hydrogen. The further
development and discussion of the atmospheric models will be presented
in future work.

\section*{Acknowledgments}

I wish to thank C. Jordan for advice and comments during all stages of 
the work presented here, G. M. Harper for many useful discussions and 
comments on this manuscript, A. D. McMurry for advice in the construction 
and use of the atmospheric models, M. Carlsson for helpful discussions and 
the referee, I. Hubeny, for useful comments and suggestions.
I also acknowledge the financial support provided 
by PPARC (D.Phil studentship).

\bibliography{Theory}

\label{lastpage}

\end{document}